\documentclass[aps,prb,twocolumn,longbibliography,superscriptaddress,amsfonts,citeautoscript,floatfix,maxnames=4,10pt]{revtex4-2}
\usepackage[utf8]{inputenc}
\usepackage[english]{babel}
\usepackage{graphicx}
\usepackage{float}
\usepackage{bm}
\usepackage{xcolor}
\usepackage{epstopdf}
\usepackage{amsmath,scalerel} 
\usepackage{amssymb}
\usepackage{epstopdf}
\usepackage[normalem]{ulem} 
\usepackage{enumerate}
\usepackage{cases}

\usepackage[urlcolor=blue,colorlinks=true,citecolor=red,linkcolor=red,pdfstartview={FitH},bookmarks=false]{hyperref}

\newcommand{\paren}[1]{\left( #1 \right)}
\newcommand{\sqbrack}[1]{\left[ #1 \right]}
\newcommand{\curlybrack}[1]{\left\{ #1 \right\}}
\newcommand{\PLL}{{\mathcal{P}^{\rm LL}}}
\newcommand{\PRR}{{\mathcal{P}^{\rm RR}}}
\newcommand{\PRL}{{\mathcal{P}^{\rm RL}}}
\newcommand{\PLR}{{\mathcal{P}^{\rm LR}}}

\graphicspath{{figs/}{./figs/}{.}}
\usepackage{lipsum}

\newcommand{\floor}[1]{\lfloor #1 \rfloor}

\begin{document}

\newcommand{\ICTMunbai}{\affiliation{Institute of Mathematical Sciences, CIT Campus, Chennai 600113, India and Homi Bhabha National Institute, Training School Complex, Anushaktinagar, Mumbai 400094, India}}

\newcommand{\JYU}{\affiliation{Department of Physics and Nanoscience Center, University of Jyväskylä, P.O. Box 35 (YFL), FI-40014 University of Jyväskylä, Finland}}

\newcommand{\LUND}{\affiliation{Solid State Physics and NanoLund, Lund University, Box 118, S-221 00 Lund, Sweden}}

\newcommand{\UPPSALA}{\affiliation{Department of Physics and Astronomy, Uppsala University, Box 516, S-751 20 Uppsala, Sweden}}

\title{Nonlocal Majorana polarization in non-Hermitian topological superconductors}

\author{Arjun S. Kumar}
\ICTMunbai

\author{Jorge Cayao}
\email[e-mail: ]{jorge.cayao@physics.uu.se}
\UPPSALA

\author{Oladunjoye A. Awoga}
\email[e-mail: ]{oladunjoye.a.awoga@jyu.fi}
\JYU

\begin{abstract}
The nonlocal Majorana polarization, defined as the product of the expectation values of the particle-hole operator at opposite halves of the system, has been shown to be a reliable topological indicator that determines the presence and quality of Majorana zero modes in Hermitian topological superconducting setups. In this work,  we extend the concept of nonlocal Majorana polarization to the non-Hermitian realm by taking into account the biorthogonal eigenstates and demonstrate its utility by exploring distinct non-Hermitian superconducting systems. In particular, we show that the Majorana polarization can distinguish between Majorana zero modes, trivial zero-energy states, and exceptional points in non-Hermitian superconductors. Also, we introduce the concept of nonlocal Majorana polarization sensitiviy for characterizing the contribution of non-Hermiticity to Majorana polarization. As a byproduct, we find that non-Hermiticity enhances Majorana zero modes robustness, a property captured by the nonlocal Majorana polarization.
\end{abstract}

\maketitle
%
%
\section{Introduction}
\label{sec.int}
Majorana zero modes (MZMs) are self-conjugated quasiparticles emerging as zero energy  edge states in topological superconductors and have triggered intense activity due to their potential for technological applications \cite{sato2016majorana,Aguadoreview17,sato2017topological,flensberg2021engineered,Marra_2022,tanaka2024theory,FukayaJPCM2025}. The realization of MZMs is well understood at the moment and requires emulating the topological phase of the Kitaev chain \cite{kitaev2001unpaired,tanaka2024theory}, e.g., using superconductor-semiconductor hybrids under a strong magnetic field \cite{Aguadoreview17,tanaka2024theory}. Currently, there are clear predicted measurable signatures of MZMs along with useful ways to characterize the topological phase \cite{flensberg2021engineered,tanaka2024theory}. This latter point is particularly important because only by identifying the topological phase  is it possible to measure true MZMs. Although   several methods have been proposed to identify the topological phase \cite{sato2016majorana,sato2017topological,tanaka2024theory}, it was shown that the nonlocal Majorana polarization (MP) is a robust real space topological indicator that signals both the topological phase and the intrinsic spatial nonlocality of MZMs \cite{PhysRevB.110.165404}. For this reason, the nonlocal MP was found to be essential to distinguish between MZMs and trivial zero energy Andreev bound states (TZESs) \cite{PhysRevB.110.165404}.  However, the usefulness of the  nonlocal MP  was only shown in the Hermitian regime, leaving unexplored   its application in  the  non-Hermitian (NH) realm.

%
%
\begin{figure}[!ht]
    \centering
        \includegraphics[width=0.48\textwidth]{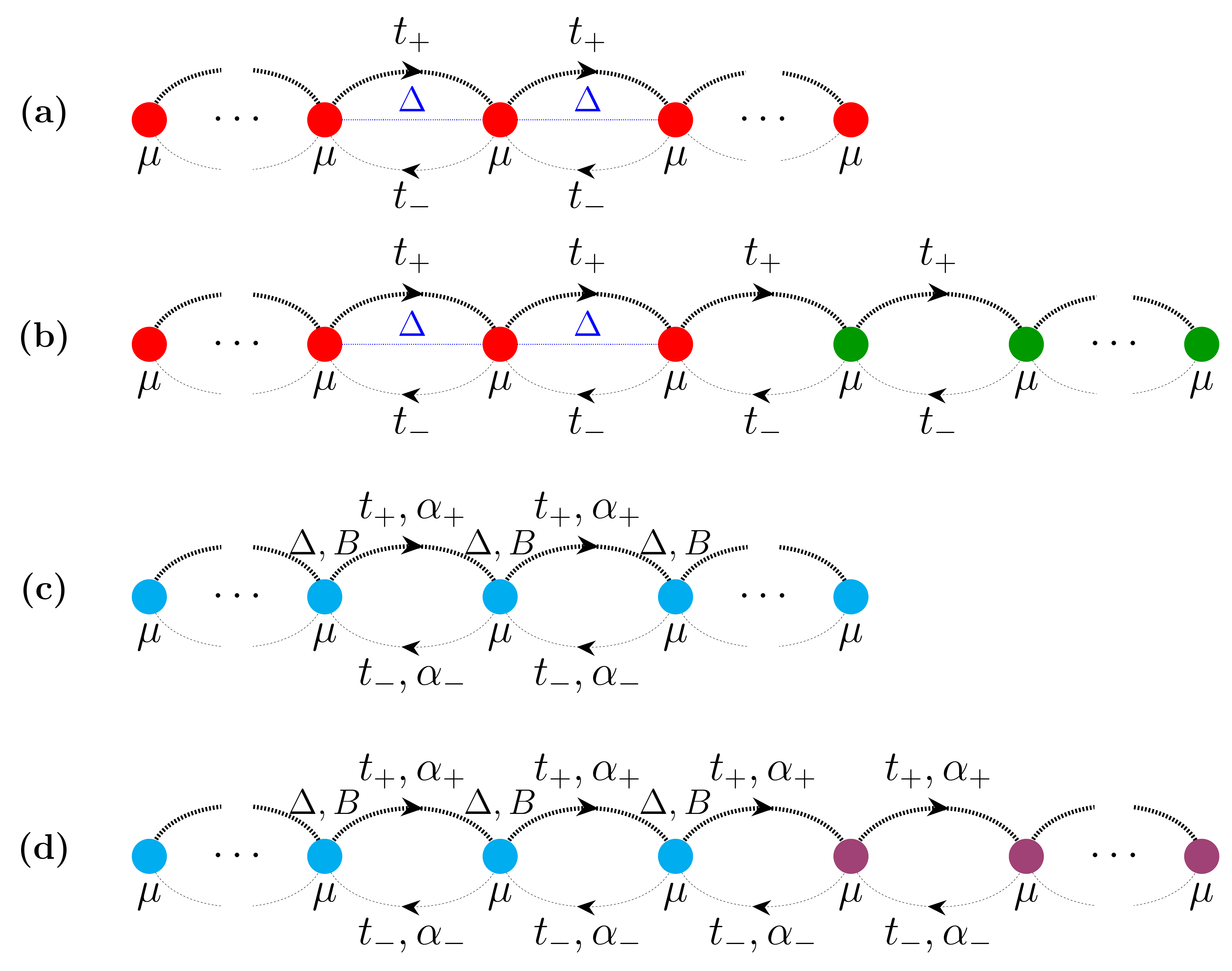}
	\caption{
 Schematics of NH superconducting chains studied in this work. (a) The Hatano-Nelson-Kitaev (HNK) chain. The red filled circles denote lattice sites with spinless fermions, $t_\pm,\,\mu$, and $\Delta$ represent nonreciprocal hopping strengths, chemical potential, and the superconducting order parameter, respectively.  (b) A superconductor-normal (SN) junction based on  the HNK chain, with the green filled circles denoting normal lattice sites.
(c) Non-Hermitian Rashba superconducting chain with Zeeman field $B$ and nonreciprocal spin-orbit strength, $\alpha_{\pm}$. The cyan filled circles represent superconducting lattice sites with spinfull fermions.
(d) SN junction of NH Rashba superconducting chain, with the brown filled circles denoting normal lattice sites.
    }
    \label{fig:Schematics}
\end{figure}

Non-Hermitian effects in superconductors  can appear under different circumstances \cite{doi:10.1080/00018732.2021.1876991,Okuma2023,doi:10.7566/JPSCP.30.011098,RevModPhys.93.015005}, such as in material junctions due to coupling to reservoirs, interactions, engineered gain and loss, and nonreciprocity. This has motivated a great deal of interest, and NH superconductors have been shown to host topological phases that otherwise would not exist \cite{JorgeEPs,PhysRevB.105.094502,PhysRevB.107.104515,PhysRevB.99.165145,PhysRevLett.123.097701,PhysRevB.103.235438,PhysRevResearch.4.L022018,PhysRevB.110.085414,cayao2023nonhermitian,li2023anomalous}; see also Refs.\,\cite{avila2019non,PhysRevB.108.L060506,shen2024nonhermitian,pino2024thermodynamics,Ohnmacht2024,cayao2024nonMulti,li2025EP,PhysRevB.111.064517,ogino2025,solow2025EP,JunjieNHDiode,cayaoSatoNH4MZMs,cayaominimalkitaev,PhysRevB.109.L161404,9jdy-b418}. Specifically, they exhibit a complex spectrum with spectral degeneracies known as exceptional points (EPs) \cite{doi:10.1080/00018732.2021.1876991}, where eigenvalues and eigenvectors coalesce, and represent a source of NH topology \cite{PhysRevX.9.041015,PhysRevX.8.031079,doi:10.7566/JPSCP.30.011098,RevModPhys.93.015005,Okuma2023}. In systems hosting MZMs, EPs have been shown to stabilize MZMs at zero real energy by considerably broadening the   parameter regime for the topological phase  \cite{JorgeEPs,PhysRevB.108.L060506,cayaoSatoNH4MZMs,cayaominimalkitaev}, while offering NH topological protection \cite{Okuma2023,cayao2023nonhermitian}. The effect of non-Hermiticity on the Majorana wavefunctions is not trivial, since it breaks down the usual bulk-boundary correspondence \cite{doi:10.1080/00018732.2021.1876991,RevModPhys.93.015005,Okuma2023},  which directly affects the spatial nonlocality of MZMs. Although there are ways to identify NH boundary states \cite{Okuma2023}, characterizing the impact of non-Hermiticity on the spatial nonlocality of MZMs remains unexplored. Therefore, it is natural to ask whether the non-local MP, which has proven useful in the Hermitian regime \cite{PhysRevB.110.165404,Kobialka2024Topological}, is capable of signaling the presence of MZM.

In this work, we extend the concept of non-local MP as a diagnostic tool to characterize zero-energy states in NH superconducting systems. In particular, we generalize the definition of the nonlocal MP by taking into account the natural biorthogonal eigenstates of NH systems, hence providing a practical way to unveil the impact of non-Hermiticity. We then utilize  this generalized biorthogonal nonlocal MP to explore the character of zero-energy states in two representative NH topological superconducting systems: the Hatano-Nelson-Kitaev (HNK), including few-site versions, and the NH Rashba superconductor. We demonstrate that the nonlocal biorthogonal MP captures the topology of NH superconductors and distinguishes between non-Majorana zero-energy states, such as EPs and TZESs, and MZMs in these systems. Furthermore,   we find that nonreciprocity enhances the  robustness of MZMs, a property that is directly unveiled by the biorthogonal nonlocal MP. Our findings offer a way to characterize zero-energy states in NH topological superconductors.

The remainder of this paper is organized as follows. In Section~\ref{sec:MHBMP}, we present the concept of nonlocal MP in NH superconductors and its generalization within a biorthogonal description. In Section~\ref{sec:HNK}, we characterize the zero-energy levels of the HNK model using the nonlocal biorthogonal MP, where we also analyze few-site HNK chains. In Section~\ref{sec:Spinful}, we use the biorthogonal MP in a non-Hermitian Rashba superconductor, and show that nonreciprocity influences the topological phase transition point.
%
\section{Nonlocal Majorana polarization in non-Hermitian systems}
\label{sec:MHBMP}
The Majorana polarization was initially  introduced and utilized as a local spatial topological indicator in Hermitian systems~\cite{sticlet.bena.12,sedlmayr.bena2015,sedlmayr.aguiarhualde.16,Bena2017,Sedlmayr.Kaladzhyan.2017.Bulk,PhysRevResearch.6.033154,Souto2024Probing,Samuelson2024Minimal,He.et.al.2025.Manipulating,karoliya2025majorana}. More recently, these ideas have generalized into the so-called nonlocal MP in Hermitian topological superconductors, which not only encodes  the topological character but also unveils the spatial nonlocality of MZMs~\cite{PhysRevB.110.165404}. Following this recent development, in this part we further generalize the definition of nonlocal MP to the non-Hermitian realm \footnote{We note that the MP has already been studied in NH systems \cite{Wang.2021.Majorana} but still as a local spatial topological indicator. This leaves the nonlocal MP in NH systems unexplored.}. 

In NH systems, having $H \neq H^\dagger$ implies that  the eigenvectors are biorthogonal~\cite{Moiseyev,Kunst.2018.Biorthogonal}. This implies that $\langle \Psi_{n^\prime}^{\rm L}| \Psi_n^{\rm R}\rangle = \delta_{n^\prime n}$, where $ |\Psi_{n}^{\rm R}\rangle\, \paren{\langle\Psi_{n}^{\rm L}|}$ is the $n$-th eigenvector of $H\,\paren{H^\dagger}$. Thus, the presence of distinct left (L) and right (R) eigenvectors in NH systems allows for alternative methods in calculating expectation values as discussed in Ref.\,\cite{Moiseyev}. In this spirit, we define the MP at each site $j$ as the expectation value of the particle-hole operator as
\begin{equation}\label{eq:MP-site}
   p_j^{a b}= \frac{\sum_{\sigma=\uparrow,\downarrow}\langle \Psi_{j\sigma}^a|\mathcal{C} | \Psi_{j\sigma}^b \rangle }{\sum_{\sigma=\uparrow,\downarrow} \langle \Psi_{j\sigma}^{a} | \Psi_{j\sigma}^{a} \rangle}, \qquad 
\end{equation}
where  $\mathcal C$ is the particle-hole operator which depends on the type of particle-hole symmetry (PHS) of the Hamiltonian, and $|\Psi_{j\sigma}^{a,b}\rangle$ is the $a,b=$L,R eigenvector of the lowest energy level (i.e., closest to zero real energy). 
Since our motivation is to characterize the spatial nonlocality of MZMs, following Ref.~\cite{PhysRevB.110.165404}, we define the MPs at the left (l) and right (r) halves of the system as 
\begin{subequations}\label{eq:MP-left-right}  
\begin{align}
P_l^{a b} &=\sum_{j =1 }^{\floor{N/2 }} p_j^{a b}, \label{eq:leftMP}
\\
P_r^{a b} & =\sum_{j =  \floor{N/2 } +1 }^N p_j^{a b}\,,\label{eq:rightMP}
\end{align}
\end{subequations}
where $\floor{ }$ in the upper (lower) bound  of the sum in the first (second) expression is the floor function. Moreover, given that the MP on the left and right sections  of the system [Eqs.\,(\ref{eq:MP-left-right})] have opposite signs, we define the nonlocal MP  as
\begin{equation}\label{eq:Nonlocal-MP}
   \mathcal P^{a b}= P_l^{a b} \paren{ P_r^{a b} }^{*}= \paren{  P_l^{a b} }^{*} P_r^{a b}, 
\end{equation}
with $\mathcal{P}^{ab}  \in \left[-1,1\right] \bigcup \curlybrack{-\infty,\infty, {\rm NaN}}$. The numerical values $\pm \infty$ (NaN) is the same analytically undefined (indeterminate). Here, $\mathcal P^{a a}$ is the nonlocal MP studied  before in the context of Hermitian systems in real space~\cite{PhysRevB.110.165404}. We refer to this component as orthogonal nonlocal MP. The component $\mathcal P^{a b},\, a \neq b$ is the nonlocal MP within a biorthogonal description of non-Hermiticity, which we refer to nonlocal biorthogonal MP. Due to its definition,   $\mathcal{P}^{a b}$ captures the topological nature of low-energy levels in superconducting NH systems as well as their spatial nonlocality. Within this picture, a low-energy level in this case could be a MZM, EP or TZES. Thus, in analogy to the Hermitian nonlocal MP~\cite{PhysRevB.110.165404}, $\mathcal{P}^{ab}=-1$ implies that the MZMs are robust nonlocal non-overlapping true zero modes, and a small deviation from $-1 <\mathcal{P}^{ab} \ll 0$ corresponds to a finite energy splitting and hence low quality MZMs. 
In the interval $-1\ll \mathcal {P}^{ab} < 1 \mathcal{}$, $\mathcal{P}^{ab}$  signals a non-MZM, e.g., a TZES. Lastly, at an EP $\mathcal{P}^{ab}$ can take the values $\pm\infty$,  since the denominator of Eq.~\eqref{eq:MP-site} vanishes at this special point. It can also evaluate to NaN when both the denominator and numerator vanishes simultaneously at the EP.

To further quantify non-Hermiticity,  we define a non-Hermitian topological indicator resulting from the determinant of the nonlocal MP in the biorthogonal basis as
\begin{equation}\label{eq:MZM-robust}
    \mathcal{R} = \det 
    \begin{bmatrix}
        \mathcal P^{\rm LL} & \mathcal P^{\rm LR}  \\
        \mathcal P^{\rm RL}  & \mathcal P^{\rm RR} 
    \end{bmatrix}	
    =
    \mathcal P^{\rm LL} \mathcal P^{\rm RR} -\mathcal P^{\rm LR} \mathcal P^{\rm RL}.
\end{equation}
Depending on the contribution of non-Hermiticity in the MP, the quantity $\mathcal R$ assumes different values, and signals distinct regimes determined by the following conditions
\begin{equation}\label{eq:R-values}
\mathcal{R}=
\begin{cases}
0,\,\quad \text{$aa=ab$ contributions (Hermitian)}\\
-1,\,\,\, \text{$ab$ contributions (non-Hermitian)}\\
(-1,0),\,\ \text{both $aa$ and $ab$ contributions} \\
\pm \infty,\,{\rm NaN}\,,\,\,\, \text{exceptional point},
\end{cases}
\end{equation}
where $a,b={\rm L,R}$, and the contributions cannot be determined at the EP.
We can understand the domain of $\mathcal R$ from Eq.~\eqref{eq:MZM-robust}. It is seen that $\mathcal{R}=0 $ for $\PLL = \PRR= \PLR=\PRL \neq 0$ when the system is Hermitian. This implies that the MP has only Hermitian  contribution~\footnote{Another scenario with $\mathcal R =0$ occurs when $\PRR = \PLL= \PLR=\PRL = 0$ for any $\Gamma$. This is a trivial solution since the superconductor does not contain  MZMs or EPs in this case.}. The value $\mathcal R=-1$ occurs when $\PRR = \PLL=0$ and $ \PLR=\PRL=-1$. In this case, it is clear that the MP has only NH (i.e., biorthognal) contribution. For the range $-1< \mathcal{R} <0$ is obtained when $\PLL = \PRR \neq  \PLR=\PRL$, which means both the orthogonal and biorthogonal components contribute to the MP. Finally,  $\mathcal R = \pm \infty,\,{\rm NaN}$ at EPs since at these special points all MP components are $\pm \infty,\,{\rm NaN}$.
The quantity $\mathcal R$ plays a role similar to sensitivity in pseudospectrum analyses~\cite{trefethen2005spectra} and turns out to be useful for assessing the robustness of MZMs \cite{PhysRevB.108.L060506}. Therefore, we refer to $\mathcal{R}$ as nonlocal  MP sensitivity.

 In what follows, we study the applicability of Eqs.~\eqref{eq:Nonlocal-MP}-\eqref{eq:R-values} to NH systems, particularly distinguishing between MZMs, EPs and TZESs. Additionally, we are interested in the impact non-Hermiticity on MZMs and how EP $\mathcal{P}^{ab}$ evolves.  When non-Hermiticity shifts the MZMs closer to zero energy from a finite energy splitting or drives $\mathcal{P}^{ab}$ toward $-1$,  we say that non-Hermiticity enhances the robustness of MZMs. We focus on two relevant models, namely the Hatano-Nelson-Kitaev and the non-Hermitian superconducting Rashba chain models.
%
%
\section{Hatano-Nelson-Kitaev model}\label{sec:HNK}
The Hatano-Nelson-Kitaev (HNK) model is based on the Kitaev chain with nonreciprocal hoppings \cite{Hatano.1996.Localization} and has been particularly important because it serves as a representative model exhibiting key non-Hermitian properties \cite{Okuma2023,ekman2026symmetry}. The HNK chain, shown schematically in Fig.~\ref{fig:Schematics}(a,b), is modeled by  
\begin{equation}\label{eq:HNK}
\begin{split}
  \mathcal   H=-&\sum_j \left[ \mu\, d_j^\dagger d_j^{}  + t_+^{}\,d_j^\dagger d_{j+1}^{} + t_-^{}\  d_{j+1}^\dagger d_j^{} \right]  +
    \\
    &
    \sum_j  \Delta\, \left[ d_{j}^\dagger d_{j+1}^\dagger - d_{j+1}^\dagger d_j^\dagger  \right]  + \textrm{ H.c.}\,,
\end{split}
\end{equation}
where  $d_j^{}$ annihilates a spinless fermion at site $j$, $\mu$ is the homogeneous chemical potential, $\Delta$ is $p$-wave pair potential and $t_\pm^{}=t\pm\Gamma$ represent the non-reciprocal hoppings. Here, $\Gamma$ encodes the non-reciprocity, while  $t$ denotes the hopping strength.

In the normal state ($\Delta=0$), which corresponds to the regular Hatano-Nelson model, Eq.\,(\ref{eq:HNK}) hosts zero-energy EPs of the order of the size of the system at $t = \pm \Gamma$ for an open chain~\cite{Hatano.1996.Localization,PhysRevX.8.031079,RevModPhys.93.015005}. We anticipate that the superconducting phase  ($\Delta\neq 0$)  modifies the condition for EPs to appear as well as lead to the emergence of MZM. We note that the HNK model in Eq.\,(\ref{eq:HNK}) has particle-hole symmetry given by $\mathcal C = \tau^x$~\footnote{This is applicable to  non-Hermitian Kitaev models with other types of non-Hermiticity~\cite{PhysRevX.9.041015,Okuma2023,cayao2023nonhermitian}}. In what follows, we use $\mathcal C$ in Eq.~\eqref{eq:MP-site} and investigate study the usefulness of MP [Eq.\,(\ref{eq:Nonlocal-MP})] as well as the nonlocal MP sensitivity [Eq.\,(\ref{eq:MZM-robust})]   to distinguish between MZMs, EPs and TZES in the HNK model [Eq.\,\ref{eq:HNK}].

%
%
\subsection{Sweet spot: Topological phase with exact zero-energy and fully nonlocal MZMs}
\label{sec:Sweep_Spot}
One of the hallmarks of topological superconductors is the emergence of MZMs states at their boundaries~\cite{kitaev2001unpaired,tanaka2024theory}. When  MZMs are perfectly localized on a single site per edge such that there is no cross-talk between them, they constitute  perfect nonlocal zero-energy states. We will refer to this regime as the sweet spot, and stress that it can occur in very long chains \cite{sato2016majorana,sato2017topological,tanaka2024theory} but also in the so-called few-site Kitaev chains \cite{Leijnse2012Parity,Sau2012,PhysRevB.110.125408,Tsintzis2022Creating,Souto2024Probing,PhysRevB.110.245144,PhysRevB.110.245412,kotetes2024nonRecifourpi,cayaominimalkitaev,vimal2025EntMKC,PhysRevB.111.115419,cayao2026nonlocal,cayao2026AEPMKC}. In this part, we focus on  the sweet spot of the HNK model [Eq.\,(\ref{eq:HNK})], Fig.~\ref{fig:Schematics}(a), at finite and vanishing chemical potentials. 

\subsubsection{Zero chemical potential}
At $\mu=0$, the sweet spot condition for MZMs is $t=\pm\sqrt{\Gamma ^2+\Delta ^2}$. This condition is global and system size independent. Using this sweet spot condition in the definitions in Eq.~\eqref{eq:MP-site}\textendash\eqref{eq:Nonlocal-MP} we obtain,
\begin{equation}\label{case1a}
    \begin{split}
   \mathcal{P}^{\rm ab}&=   \left\{
    \begin{aligned} 
 -\frac{\Delta ^2}{\Gamma ^2+\Delta ^2},\quad a =b
 \\
   -1, \quad \qquad a\neq b
\end{aligned} 
\right.
\\
    \mathcal{R} &=  -1+ \sqbrack{\frac{\Delta^2 }{\Gamma ^2+\Delta ^2}}^2.
    \end{split}
\end{equation}
These general results demonstrate that fully nonlocal MZMs localized at each edge are characterized  by  $\mathcal{P}^{ab}=-1$ in the NH case when $\Gamma\neq0$, provided $a\neq b$.  In the Hermitian regime, when $\Gamma=0$, both nonlocal MPs from Eqs.\,(\ref{case1a}) become equal to $-1$, namely,  $\mathcal{P}^{aa}\paren{\Gamma\rightarrow 0}=\mathcal{P}^{ab}=-1$. We note that the impact of non-Hermiticity on MZMs increases with  $\Gamma$, and the MZMs  attain pure non-Hermiticity (or maximum stability) for $\Gamma \gg \Delta$, where $\mathcal{R}=-1$.

%
\begin{figure*}[!t]
    \centering
        \includegraphics[width=1\textwidth]{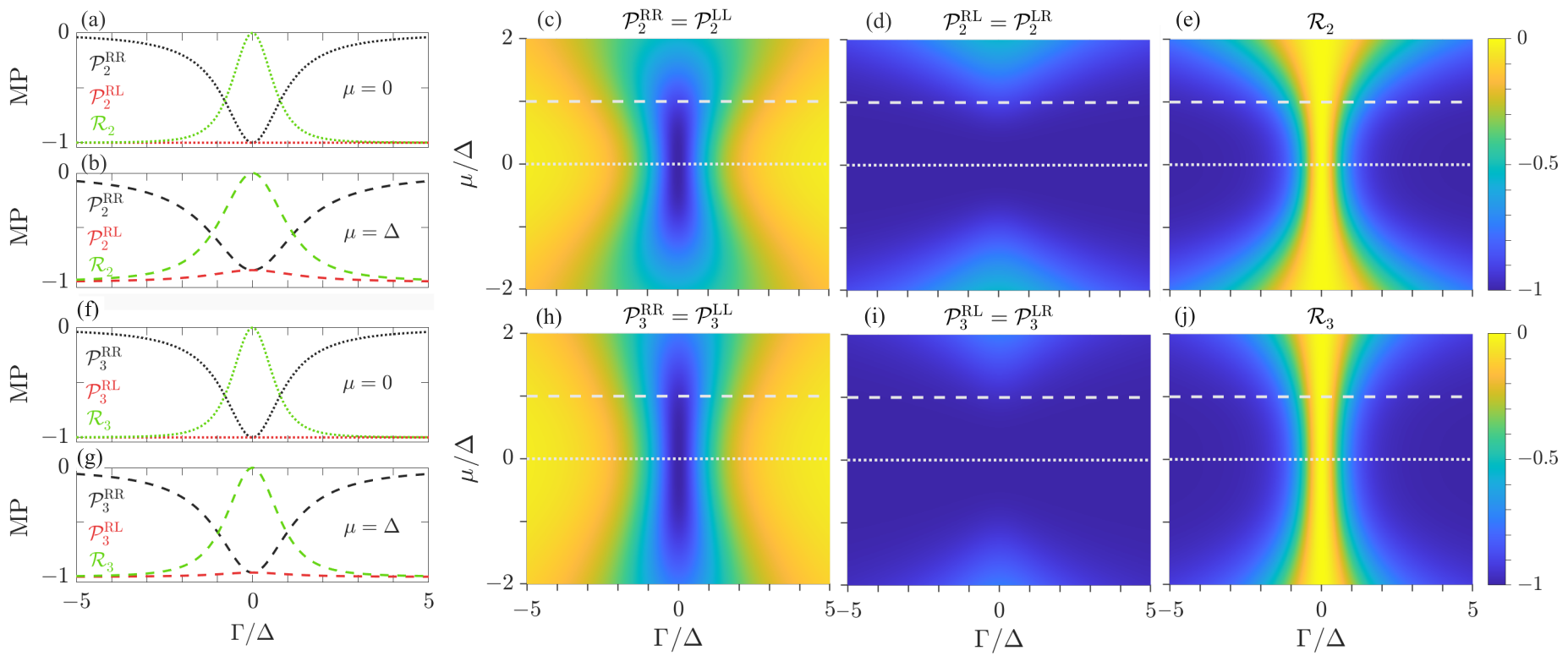}
	\caption{(a-e) Nonlocal MP biothorgonal indicators ($\mathcal{P}_{\rm N}^{\rm ab}$ and $\mathcal R_{\rm N}^{}$, with $a,b={\rm L,R}$) for a two-site HNK chain, while (f-j) for a  three-site chain; here $N$ indicates the number of system sites.  (a,b) $\mathcal R_2^{}$ and  $\mathcal P_2^{ab}$ as functions of $\Gamma$ at $\mu =0$ (a) and $\mu=\Delta$ (b). (c-e) $\mathcal P_2^{\rm aa}$ (c), $\mathcal P_2^{\rm ab}$ (d), and $\mathcal R_2^{}$ (e) as functions of $\Gamma$ and $\mu$. The dotted and dashed lines in (c-e) indicate the line cuts at $\mu=0$ and $\mu=\Delta$ shown in (a) and (b), respectively. (f-j) The same as in (a-e) but for a three-site HNK chain.}
	\label{fig:FiniteSizeMP}
\end{figure*}
%
\subsubsection{Finite chemical potential: Minimal chains}
At finite chemical potential, $\mu \neq 0$, the condition for the emergence of zero-energy becomes sensitive to the size of the system \cite{PhysRevB.108.L060506}. As the system size grows, the analytical tractability becomes  challenging. Therefore, we focus on   experimentally accessible minimal chains \cite{dvir2023realization,Haaf2024,Zatelli_2024,bordin2025enhanced,Kulesh.2025.Flux}.
%

\emph{Two-site HNK chain. } In this model, Eq.~\eqref{eq:HNK} consists of   two sites. In the Hermitian regime ($\Gamma =0$), this minimal Kitaev chain has been shown to host poor man's Majorana modes \cite{Leijnse2012Parity,Sau2012,PhysRevB.110.125408,Tsintzis2022Creating,Souto2024Probing,PhysRevB.110.245144,PhysRevB.110.245412,kotetes2024nonRecifourpi,cayaominimalkitaev,vimal2025EntMKC,PhysRevB.111.115419,cayao2026nonlocal,cayao2026AEPMKC}, which, although symmetry protected \cite{cayao2026nonlocal}, are not tied to  topology. At the moment, there also exists  experimental evidence of the realization of this minimal chain ~\cite{dvir2023realization,Kulesh.2025.Flux}. 

In the NH case, the sweep spot condition is given by $t=\pm\sqrt{\Gamma ^2+\Delta ^2+\mu ^2}$. The nonlocal MP components and sensitivity at the sweet spot are obtained as
\begin{equation}\label{eq:MP-2sites}
 \begin{split}
  \mathcal{P}_2^{\rm ab} &=   \left\{
    \begin{aligned} 
 -\frac{1 + \paren{\frac{\mu}{\Delta}}^2}{\paren{\frac{\Gamma}{\Delta}}^2 + \sqbrack{1+\frac{1}{2}\paren{\frac{\mu}{\Delta}}^2}^2}\,,\quad  a =b 
 \\
   -\frac{
        1 + \paren{\frac{\mu}{\Delta}}^2 + \paren{\frac{\Gamma}{\Delta}}^2 
        }{
        \paren{\frac{\Gamma}{\Delta}}^2 + \sqbrack{1+\frac{1}{2}\paren{\frac{\mu}{\Delta}}^2}^2
        }\,, \quad a\neq b
\end{aligned} 
\right.
      \\
    \mathcal{R}_2 &=     -\frac{
       \paren{\frac{\Gamma}{\Delta}}^2 \sqbrack{ 2 + 2\paren{\frac{\mu}{\Delta}}^2 + \paren{\frac{\Gamma}{\Delta}}^2 }
        }{
        \sqbrack{\paren{\frac{\Gamma}{\Delta}}^2 + \paren{1+ \frac{1}{2}\paren{\frac{\mu}{\Delta}}^2}^2}^2
        },
    \end{split}
    \end{equation}
 where the subscript 2 indicates that these expressions correspond to a two-site HNK minimal chain.  We note that the expressions in Eq.~\eqref{eq:MP-2sites} reduce to the corresponding global results Eq.~\eqref{case1a} at $\mu=0$. By a direct inspection, we identify that, as $\Gamma/\Delta \gg \mu/\Delta$,  $\{\mathcal{P}_2^{ab},\, \mathcal{R}_2^{}\}\rightarrow -1$ and $\mathcal{P}_2^{aa}\rightarrow 0$, unveiling  that non-Hermiticity enhances the  robustness of MZMs. 

For a better understanding of the above discussion, in Fig.~\ref{fig:FiniteSizeMP}(a-e) we present the nonlocal biorthogonal MP and the sensitivity indicator [Eqs.\,(\ref{eq:MP-2sites})] as functions of $\Gamma$ and $\mu$. In Hermitian systems, it is known that $\mu \neq 0$ reduces the protection of MZMs leading low quality Majoranas even at the sweet spot~\cite{Leijnse2012Parity,PhysRevB.110.165404}. Interestingly, Fig.~\ref{fig:FiniteSizeMP}(a-e) reveals that the non-Hermitian regime has an intriguing response to $\mu$. At $\mu =0$ in Fig.~\ref{fig:FiniteSizeMP}(a), the  nonlocal biorthogonal MP $\mathcal P_2^{\rm RL}=-1$ and independent of $\Gamma$, while the nonlocal orthogonal MP  $\mathcal P_2^{\rm RR}$ increases from $-1$ at $\Gamma=0$ to $0$ at $|\Gamma|\gg \Delta$. Furthermore, the nonlocal MP sensitivity 
 $\mathcal R_2^{\rm }$ develops a behavior that is opposite  to $\mathcal P_2^{\rm RR}$, reaching zero at $\Gamma=0$ (pure Hermitian) but achieving values of $-1$ (pure NH) for $|\Gamma|/\Delta>2$. Notably, these results show that nonlocal MP indicators indeed acquire the values of those indicators for a  vanishing chemical potential scenario [Eq.~\eqref{case1a}].  At finite $\mu \neq 0$ in Fig.~\ref{fig:FiniteSizeMP}(b),  the quality of MZMs is reduced at the sweet spot for small $|\Gamma |/\Delta <1$, which is revealed by noting that the nonlocal MPs not achieving $-1$. By increasing $\Gamma$, the values of the nonlocal MPs and MP sensitivity reach $-1$, namely,  $\mathcal P_2^{\rm RL}=\mathcal R_2^{} =-1$. Thus, non-Hermiticity via $\Gamma\neq0$ can restore the robustness of MZMs at  $\mu \neq 0$, thus giving access to high quality MZMs. 

For completeness, in Fig.~\ref{fig:FiniteSizeMP}(c-e) we present the nonlocal MP indicators as a function of $\Gamma$ and $\mu$, hence supporting the above results for arbitrary values of $\mu$. First, we observe that the nonlocal orthogonal MPs, $\mathcal{P}_2^{\rm aa}$, exhibit values equal to $-1$ around $\Gamma=0$, which remain roughly constant as $\mu$ increases up to $\mu=\Delta$ but then decay for $\mu\gg\Delta$. For large values of non-Hermiticity ($|\Gamma|>\Delta$) $\mathcal{P}_2^{\rm aa}$ decays algebraically to zero, indicating that it fails to capture the MZMs in the system. Therefore, $\mathcal{P}_2^{\rm aa}$ is not a reliable topological indicator for NH superconductors. In the case of the nonlocal biorthogonal MP $\mathcal{P}_2^{\rm RL (LR)}$ in Fig.~\ref{fig:FiniteSizeMP}(d), which is entirely of non-Hermitian origin, it achieves values close to $-1$ for $|\mu|\leq \Delta$, but larger chemical potential values unavoidably lead to a reduction of $\mathcal{P}_2^{\rm RL (LR)}$. The decay of these nonlocal MP indicators can be understood from Eqs.\,(\ref{eq:MP-2sites}), which, for $\mu\gg |\Gamma|,\, \Delta$, lead to $\mathcal{P}_2^{\rm RL} = \mathcal{P}_2^{\rm RR} \approx-\paren{\Delta/\mu}^2$. In this regime, non-Hermiticity is rather tiny, and the values of the MP indicators can be compared with previous Hermitian results reported in the Appendix of Ref.~\cite{PhysRevB.110.165404}. Furthermore, for $|\Gamma|>\Delta$, the MP sensitivity $\mathcal{R}_2^{}$  shown in Fig.~\ref{fig:FiniteSizeMP}(e) becomes of the order of $-1$, implying that it is entirely  non-Hermitian: in this case, having $\mathcal{R}_2^{}=-1$ means that biorthogonal MP, is the main contributor [Eq.\,(\ref{eq:MZM-robust})]. As before, this behavior of $\mathcal{R}_2^{}$ can be directly seen by obtaining it from Eqs.\,(\ref{eq:MP-2sites}) for large chemical potential, which gives $\mathcal{R}_2^{} \approx- 2\paren{2\Gamma/\mu}^2$: hence, $\mathcal{R}_2^{} \rightarrow 0$ as $\Gamma/\mu\rightarrow 0$, consistent with the observation in  Fig.~\ref{fig:FiniteSizeMP}(e). This analysis is also supported by using Eqs.\,(\ref{eq:MP-2sites}) to obtain   expressions for a large chemical potential but keeping  the lowest contribution in $\Gamma$: in the limit $\mu\gg |\Gamma| \gg \Delta$, we get $ \mathcal{P}_2^{\rm RR} \approx 0$, $\mathcal{P}_2^{\rm RL} \approx-\paren{2\Delta/\mu}^2\big[1+ \paren{\Gamma/\mu}^2\big]$, and $\mathcal{R}_2^{} \approx -\paren{2\Gamma/\mu}^2\big[2+ \paren{\Gamma/\mu}^2\big]$, all in agreement with the behavior shown in Figs.~\ref{fig:FiniteSizeMP}(c-e). The behavior of   $\mathcal R_2^{}$ and $\mathcal{P}_2^{\rm RL}$, which do not have a Hermitian analogue, clearly shows  that non-Hermiticity enhances  the nonlocal biorthogonal MP in comparison to the Hermitian scenario, $ \mathcal{P}_2^{\rm RR} \approx 0$. This can be directly interpreted as the enhancement of MZM robustness by non-Hermiticity.

%
%
%
\emph{Three-site HNK chain.} We now turn our attention to another relevant minimal HNK model consisting of three sites. Its Hermitian version is  currently attracting great attention from both theoretical~\cite{Dourado.2025.Majorana,2r8x-9d9m,vimal2025EntMKC} and experimental~\cite{bordin2025enhanced,ten_Haaf_2025,bordin2026probing} perspectives, while its non-Hermitian regime induced by nonreciprocity via $\Gamma$ still remains seldom studied. 
In the NH phase, the sweet spot condition of the three-site HNK model is given by 
$t= \pm \sqrt{\Gamma ^2+\Delta ^2+(\mu^2/2)}$ and, in this regime, we obtain  the nonlocal MPs as well as the MP sensitivity
    \begin{equation}
    \begin{split}
         \mathcal{P}_3^{\rm ab}& =   \left\{
    \begin{aligned}\label{eq:MP-3sites}
 -\frac{
      1 + \frac{1}{2}\paren{\frac{\mu}{\Delta}}^2
      }{
     \paren{\frac{\Gamma}{\Delta}}^2  + \sqbrack{ 1  +  \paren{\frac{\mu}{2\Delta}}^2}^2
      } , \quad a =b
 \\
   -\frac{
        1 + \frac{1}{2}\paren{\frac{\mu}{\Delta}}^2 + \paren{\frac{\Gamma}{\Delta}}^2 
        }{
         \paren{\frac{\Gamma}{\Delta}}^2  + \sqbrack{ 1  +  \paren{\frac{\mu}{2\Delta}}^2}^2
         }, \quad a\neq b 
\end{aligned} 
\right.
       \\
    \mathcal{R}_3 &=     -\frac{
       \paren{\frac{\Gamma}{\Delta}}^2 \sqbrack{ 2 + \paren{\frac{\mu}{\Delta}}^2 + \paren{\frac{\Gamma}{\Delta}}^2 }
        }{
        \sqbrack{ \paren{\frac{\Gamma}{\Delta}}^2  + \sqbrack{ 1  +  \paren{\frac{\mu}{2\Delta}}^2}^2 }^2
        }\,,
    \end{split}
    \end{equation}
where subscript 3 indicates that these expressions correspond to a three-site HNK chain. Similarly to the two-site scenario discussed in the previous subsection, Eqs.~\eqref{eq:MP-3sites} reduce to the global results given by Eqs.~\eqref{case1a} at $\mu=0$, but  provide more information when $\mu\neq0$.

The first observation of  Eqs.\,(\ref{eq:MP-3sites}) is that they exhibit a dependence on $\mu$, $\Delta$, and $\Gamma$ that is similar to those of Eq.\,(\ref{eq:MP-2sites}) for  the two-site HNK chain; they only differ in the coefficients of $(\mu/\Delta)^2$. This can indeed be seen in Fig.~\ref{fig:FiniteSizeMP}(f-j), where we plot the MP indicators given by Eqs.\,(\ref{eq:MP-3sites}), which reflect qualitatively the same behavior as those of two-site chain [Fig.~\ref{fig:FiniteSizeMP}(a-e)]. The only appreciable difference occurs at $\mu \neq 0$: $\mathcal{P}_{3}^{\rm RL}$ in Fig.~\ref{fig:FiniteSizeMP}(g) practically remains at $-1$ for all $\Gamma$ and $\mathcal{P}_{3}^{\rm RR}$ reaches $-1$ only at $\Gamma=0$,
while the MP indicators for the two-site chain differ from $-1$ in the same regime of nonreciprocity  [Fig.~\ref{fig:FiniteSizeMP}(b)]. Note that having  $\mathcal{R}_{3}^{\rm }\rightarrow-1$ means that there are only NH contributions, an instance that is easier to achieve in the three-site chain. Further insights on these results can be obtained by inspecting the nonlocal MPs given by Eqs.\,(\ref{eq:MP-3sites}) in limit $\mu\gg |\Gamma|,\, \Delta$: here, we obtain $\mathcal{P}_3^{\rm RL} = \mathcal{P}_3^{\rm RR}  \approx-2\paren{\Delta/\mu}^2 = 2\mathcal{P}_2^{\rm RL}$ and $\mathcal{R}_3^{}\approx -\paren{4\Gamma/\mu}^2= 2 \mathcal{R}_2^{}$. Hence, the three-site HNK chain clearly showing  a factor of two improvement in the  robustness of MZMs  compared to the two-site chain. This is also supported by Fig.~\ref{fig:FiniteSizeMP}(h-j), which shows the nonlocal MP indicators for the three-site chain as functions of $\Gamma$ and $\mu$. By comparing with the two-site results [Fig.~\ref{fig:FiniteSizeMP}(c-e)], it is evident that the  three-site chain MP indicators show improved values. 
This view can by further supported  by analyzing Eq.~\eqref{eq:MP-3sites} in the limit $\mu\gg |\Gamma| \gg \Delta$, which then reduce to $\mathcal{P}_3^{\rm RR}=\mathcal{P}_2^{\rm RR}\approx 0,\,\mathcal{P}_3^{\rm RL} \approx-2\paren{2\Delta/\mu}^2\big[1+2 \paren{\Gamma/\mu}^2\big] < \mathcal{P}_2^{\rm RL}$ and $\mathcal{R}_3^{} \approx - \paren{4\Gamma/\mu}^2\big[1+ \paren{\Gamma/\mu}^2\big]<\mathcal{R}_2^{}$. The last two expressions clearly show that MZMs are better protected in the three-site chain compared to the two-site chain since $\mathcal{P}_3^{\rm RL} < \mathcal{P}_2^{\rm RL}$ implies that  MZMs in the former are better localized.

Following the above analysis, we could  explore four-site and larger chains in principle, but the analytical calculations offer limited insights \footnote{For example, the four-site model has two sweet spot conditions, namely
$t= \pm \sqrt{\Gamma ^2+\Delta ^2+\frac{1}{2} \left(\sqrt{5}+3\right) \mu ^2}$ and $t= \pm \sqrt{\Gamma ^2+\Delta ^2+\frac{1}{2} \left(\sqrt{5}-3\right) \mu ^2}$. Solving Eqs.~\eqref{eq:MP-site}-\eqref{eq:MZM-robust} for either of these sweet spots yields expressions for $\mathcal{P}^{ab}$ and $\mathcal{R}$ which are complicated and do not provide intuitive insights of the behavior of the nonlocal MPs.}. However, we find that increasing the  size of the system does not change the qualitative behavior of the nonlocal MP indicators, but it enlarges the parameter space with $\mathcal{P}^{\rm RL} \approx-1$, indicating the formation of high quality MZMs, in agreement with earlier studies~\cite{PhysRevB.95.184511,PhysRevLett.123.117001,Samuelson2024Minimal}. Thus, our results show that in NH systems the nonlocal biorthogonal MP ( $\mathcal{P}^{\rm RL}=\mathcal{P}^{\rm LR}$) characterizes MZMs as well as distinguishes MZMs and non-MZM zero-energy states. Furthermore, the sensitivity 
($\mathcal{R}$) captures the contribution of non-Hermiticity to the MP and also characterizes the robustness of MZMs. We find that increasing $\Gamma$ enhances both   $\mathcal{P}^{\rm RL}$ and $\mathcal{R}$ either for vanishing or finite chemical potential, which implies that non-Hemiticity enhances MZMs robustness.
%
%
%
\subsection{Away from the sweet spot: Finite energies and topological phase transitions} 
\label{sec:TopTrans}
\begin{figure}[!t]
    \centering
        \includegraphics[width=0.48\textwidth]{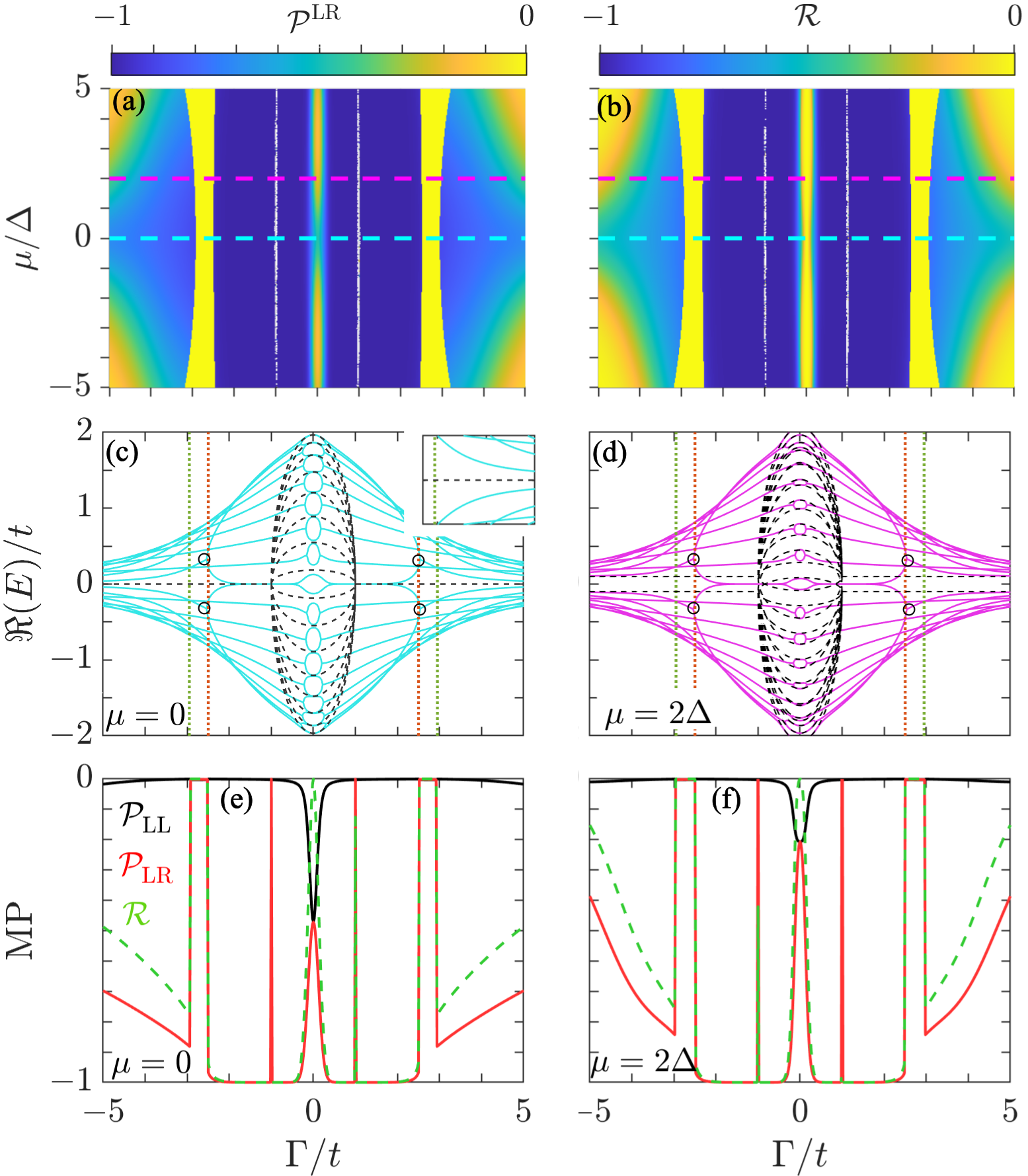}
	\caption{Nonlocal MP indicators and  spectral features of a long HNK chain away from the sweet spot. (a)  $\PLR$ and (b) $\mathcal{R}$ in $\mu-\Gamma$ space.
     (c,d) $\Re (E)$ as a function of $\Gamma$ with colors corresponding to dashed lines in (a). Solid (dashed) corresponds to $\Delta\neq0$ ($\Delta = 0$). The dotted vertical lines represent transition points, with red (green) denoting EP (level crossing) point. The red circles highlight the EPs at large $\Gamma$. Inset in (c): zoom-in of the spectrum around zero energy. (e,f) Nonlocal MP indicators as a function of $\Gamma$ for the lowest energy levels in (c,d)[ (cyan, magenta) line in (a,b)]. Here $\Delta/t = 0.05$ and $N=16$ sites. }
	\label{fig:HNK_Chain}
\end{figure}

Thus far, we have focused on the topological phase at the sweet spot, where  MZMs are exactly at zero energy and perfectly localized on a single site of few-site HNK chains. In practice, however, achieving exact zero energy is  challenging in the Hermitian regime and MZMs typically acquire a finite energy; this is not strictly restricted to short systems, but also occurs in long Hermitian topological superconductors \cite{PhysRevB.86.180503,PhysRevB.91.024514,PhysRevB.96.205425}. Moreover, it is also known that the Hermitian regime of Majorana devices, aimed at realizing the Kitaev chain, can also host TZESs below the topological phase transition and thus without any relation to topology  \cite{PhysRevB.91.024514,PhysRevB.98.245407,PhysRevB.104.L020501,PhysRevLett.123.117001,PhysRevB.105.144509,PhysRevB.107.184519,PhysRevB.104.134507,PhysRevLett.130.207001,fksg-x8pr,prada2019andreev}. It is therefore of interest to assess the impact of non-Hermiticity ($\Gamma\neq0$) on the regime away from the sweet spot in longer chains and characterize it by the biorthogonal nonlocal MP, $\mathcal{P}^{ab}$, and nonlocal MP sensitivity, $\mathcal{R}$, introduced in Section \ref{sec:MHBMP}. For this purpose, we numerically analyze all the eigenvalues of Eq.~\eqref{eq:HNK} and the MP indicators obtained from the  energy state closest to zero real energy. Thus, in Fig.~\ref{fig:HNK_Chain} we present $\mathcal{P}^{ab}$ and $\mathcal{R}$ as functions of $\Gamma$ and $\mu$ away from the sweet spot in a HNK chain with $N=16$ sites. For completeness, we also show the real parts of the spectrum, $\Re (E)$, without ($\Delta=0$) and with ($\Delta\neq0$) superconductivity. 
%
%

Figure~\ref{fig:HNK_Chain}(a) and~\ref{fig:HNK_Chain}(b) show the MP indicators, $\PLR$ and $\mathcal{R}$, respectively. Noticeably, in the regime $|\mu|\le \Delta$ the system undergoes a re-entrant topological phase transition with $\Gamma$: the system first transitions from topological phase $(\PLR,\, \mathcal{R}=-1)$ to trivial phase $(\PLR,\, \mathcal{R}=0)$ and re-enters the topological phase $(-1 < \PLR,\, \mathcal{R} <0)$ with further increase in $\Gamma$. After the re-entering the topological phase the quality of the MZMs reduces, $-1<\PLR,\, \mathcal{R} \ll 0$. Still, Fig.~\ref{fig:HNK_Chain}(b) shows that the non-Hermitian topological phase is still favorable over the Hermitian counterpart in this parameter regime. We note that in the vicinity of $|\Gamma/t|=1$, $\PLR,\, \mathcal{R} \rightarrow \infty$ or NaN numerically (but undefined or indeterminate analytically) leading to the thin white regions in Fig.~\ref{fig:HNK_Chain}(a,b). The reason is that the normal state EPs at $|\Gamma/t|=1$ persist in the presence of superconductivity. This result shows that the EPs do not transition to MZMs despite the presence of MZMs ingredients, contrary to the conclusion of earlier works~\cite{PhysRevB.108.L060506}. Although the evolution of EPs in the presence of superconductivity is an interesting topic, it is beyond the scope of the current study as we are mainly interested in the lowest energy levels, but the topology of EPs can be studied using biorthogonal polarization~\cite{Kunst.2018.Biorthogonal,RevModPhys.93.015005}. Note that the biorthogonal polarization is purely electronic while the biorthogonal MP is electron-hole in nature.

To understand the phase diagram it is instructive to study the lowest energy states in the chain, particularly evolution with $\Gamma$. Figures~\ref{fig:HNK_Chain}(c) and~\ref{fig:HNK_Chain}(e) show the $\Re(E)$ and the corresponding MPs at $\mu=0$ (dashed cyan line in Fig.~\ref{fig:HNK_Chain}(a) and~\ref{fig:HNK_Chain}(b)), respectively. The EPs of the normal state, existing at $\paren{E=0,|\Gamma|/t=1}$ (dashed black) are modified by superconductivity: the new EPs arise from electron-hole levels merging at finite energy which depends on $\Gamma$, with all but one of the EPs occurring at $|\Gamma|/t <1$. The last EP moves to the topological phase transition point [black circles in Fig.~\ref{fig:HNK_Chain}(c)].  At $\Gamma=0$, the eigenvalues are real and the MZMs have finite energy due to finite size effect. Here,  $-1<\PLR=\PLL \ll 0$ and the MZMs are of low quality. As $\Gamma$ increases, the chain becomes non-Hermitian and enhances the robustness of MZMs to high quality ones such that they attain zero energy, leading to $\PLR=-1,\,\PLL =0$, when the system is purely non-hermitian, i.e., $\mathcal{R}=-1$. We note that $\PLR$ jumps around $\Gamma/t=1$~\footnote{The value numerically jumps to infinity  but we have shown it as a jump to zero in Fig.~\ref{fig:Schematics}(c,d) for clarity.}, implying that the zero energy states around this point are not MZMs but EP due to unidirectional hopping.  With further increase in $\Gamma$, the system becomes completely gapped and there are no MZMs at the ends of the chain i.e. the HNK chain transitions to trivial phase [see dotted vertical lines in Fig.~\ref{fig:HNK_Chain}(c)].  Here $\Re(E)\neq 0$ and $\PLR = \PLL = \mathcal{R}=0$. Further increasing $\Gamma$ the chain re-enters the topological phase but the MZMs are of low quality or robustness, $-1< \PLR \ll 0$, which further reduces with $\Gamma$. The reduced robustness of the MZMs in this regime arises from the small energy gap (called topological gap) separating them from the nearest excited states [see inset in Fig.~\ref{fig:HNK_Chain}(c)]. Figs.~\ref{fig:HNK_Chain}(d) and~\ref{fig:HNK_Chain}(f) show that while finite $\mu$ does not change the qualitative picture discussed here in the regime $|\Gamma|<t$, it detrimental  impact on the re-entrant regime. As seen in Fig.~\ref{fig:HNK_Chain}(d) the re-entrant topological phase is destroyed by $\mu$. This is consistent with Fig.~\ref{fig:HNK_Chain}(a,b) which shows that $\PLR=0$ and $\mathcal{R} \approx 0$ when $|\mu| \ge \Delta$ at $|\Gamma|\gg t$. It is worth noting that, while  NH systems are  sensitive to system size, the results presented above remain unchanged by increasing the length of the chain; the only change is that the values of $\Gamma$ at which the topological and re-entrant  transitions occur changes with chain length, provided all other parameters fixed. We can thus conclude that the nonlocal biorthogonal MP ($\PLR$) and the sensitivity ($\mathcal{R}$)  identify MZMs and also distinguish them from EPs despite being away from the sweet spot. These results agree with the sweet spot analysis in Section~\ref{sec:Sweep_Spot}, and confirm that $\PLR$ and $\mathcal{R}$ are good indicators of the topological phase in real space in NH topological superconductors.
%
%
%
%
\begin{figure}[!ht]
    \centering
        \includegraphics[width=0.3\textwidth]{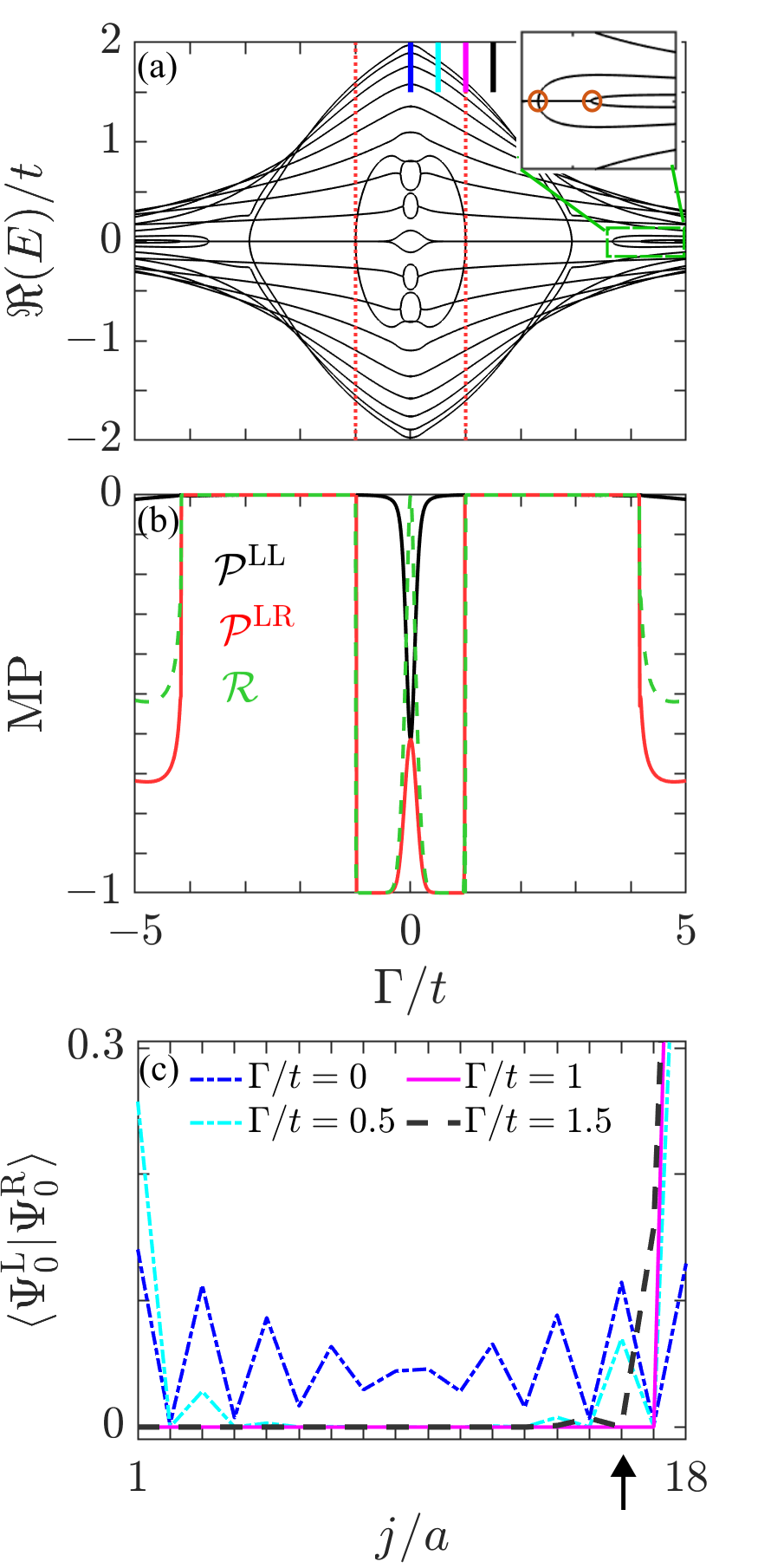}
	\caption{Spectrum and nonlocal MP indicators in HNK junction. (a) Real part of the spectrum, $\Re (E)$, with dotted vertical lines denoting the EPs at $|\Gamma|/=t$. Inset: zoom-in of the spectrum around zero energy at large $\Gamma$.(b) Nonlocal MP indicators as functions of $\Gamma$.  (c) Probability density as a function of site for several $\Gamma/t$. Colors correspond to the bars in (a) while the arrowhead indicates the boundary between N and S regions. Here, $N_{\rm S}^{}=16$, $N_{\rm n}^{}=2$, $\mu=0$, all others parameters are same as Fig.~\ref{fig:HNK_Chain}.}
	\label{fig:HNK_NS}
\end{figure}
\subsection{Superconductor-Normal metal junctions in HNK chains }
In typical experimental setups, the ends of the superconductor are normal (i.e., nonsuperconducting) to allow contact with probes~\cite{Deng16,dvir2023realization,bordin2025enhanced}. Due to spatial confinement, this normal region can host accidental TZESs~ \cite{PhysRevB.91.024514,PhysRevB.98.245407,PhysRevB.104.L020501,PhysRevLett.123.117001,PhysRevB.105.144509,PhysRevB.107.184519,PhysRevB.104.134507,PhysRevLett.130.207001,fksg-x8pr,prada2019andreev}. It is therefore naturally to expect that such TZESs can also emerge in NH superconductors, in addition to EPs. In this section, we will test the ability of nonlocal biorthogonal MP, $\mathcal{P}^{ab}$, to distinguish between these TZESs, EPs and MZMs.

For this purpose, we consider an SN junction made of the HNK chain modelled by Eq.~\eqref{eq:HNK} and depicted in Fig.~\ref{fig:Schematics}(b): here, the superconducting order parameter is finite in the S region but vanishes in the N region, i.e., $\Delta>0\, \forall\, j \in N_{\rm S}$ and $\Delta=0,\,  N_{\rm S} < j \le N =N_{\rm S} +N_{\rm n}$, $N_{\rm S}\, \paren{N_{\rm n}}$ denotes the number of superconducting (normal) sites. In Fig.~\ref{fig:HNK_NS}, we plot the spectrum (a), MP (b) and probability density (c) for $\mu=0, \, N_{\rm S}=16,\, N_{\rm n}=2$. 
For $|\Gamma|/t <1$, the junction is that of a topological superconductor and has the same low energy features and MP as the single HNK investigated in Section~\ref{sec:TopTrans}, see Fig.~\ref{fig:HNK_NS}(a,b). At $\Gamma=0$ in the Hermitian regime,  Majorana states have finite energy due to finite size effects leading to poor localization, i.e., low quality MZMs [see blue in Fig.~\ref{fig:HNK_NS}(c)]. In the non-Hermitian regime, $\Gamma\neq0$ increases the robustness of  MZMs, which then achieve zero-energy and improved localization  at the ends of the system albeit the MZM at the right end leaks into the N part [see cyan in Fig.~\ref{fig:HNK_NS}(c)]. At $|\Gamma|/t=1$, the chain has an EP, as seen in Fig.~\ref{fig:HNK_NS}(c) (magenta) the zero-energy state resides at a single site with probability $1$ (maximum value has been cut-off in the figure for clarity). Remarkably, at this point $\PLR$ jumps to zero [see Fig.~\ref{fig:HNK_NS}(b)]. Beyond the point, $|\Gamma|/t>1$, $\PLR=\PRR=\mathcal R =0$ despite the presence of zero-energy level. In this region, the zero-energy level lies in the normal region [see black in Fig.~\ref{fig:HNK_NS}(c)], showing that it is a TZES.  These results demonstrate that $\PLR$ provides an effective way to distinguish between MZMs and spatially confined non-Majorana states. Up to this point, our analysis has focused on the spinless HNK chain; we next apply it to the spinfull regime.
%
\section{Non-Hermitian Rashba superconducting chains}
\label{sec:Spinful}
We now utilize the nonlocal MP indicators to assess the effect of non-Hermiticity on a NH Rashba superconductors, shown schematically in Fig.~\ref{fig:Schematics}(c,d). We particularly focus on non-Hermiticity due to nonreciprocal hopping and Rashba spin-orbit coupling. The non-Hermitian Rashba superconductor is modeled by
\begin{equation}\label{eq:SpinfulH}
    \begin{split}
\mathcal{H} & = 
\sum_{\substack{j=1\\ \sigma\sigma'}}^{M} 
 d_{j\sigma}^\dagger \lbrace 
\left[ \varepsilon\,\delta_{\sigma\sigma'}
+ B\sigma^x_{\sigma\sigma'} 
 \right] 
d_{j\sigma'}   + \left[ i\Delta \sigma^y_{\sigma\sigma'}\right]
d_{j\sigma'}^\dagger \rbrace
\\
&  + \sum_{j=1,\sigma\sigma'}^{M-1} 
d_{j\sigma}^\dagger 
\left[ -t_+\,\delta_{\sigma\sigma'} 
+ \alpha_+\,\sigma^y_{\sigma\sigma'} \right] 
d_{j+1,\sigma'}\ +  
\\
& \sum_{j=1,\sigma\sigma'}^{M-1} 
d_{j+1,\sigma}^\dagger 
\left[ -t_-\,\delta_{\sigma\sigma'} 
- \alpha_-\,\sigma^y_{\sigma\sigma'} \right] 
d_{j,\sigma'} + \text{H.c.},
\end{split}
\end{equation}
where  $d_{j\sigma}^{} $ creates and annihilates a particle of spin $\sigma$ at site j. Here $\varepsilon = 2t - \mu$ is the onsite energy with $\mu$ being the chemical potential, $B$ is the Zeeman field, $\Delta$ is the superconducting order parameter with $s$-wave symmetry, $t_\pm^{}=t\pm\Gamma$, where $\Gamma$ encodes the non-reciprocity and $t$ denotes the hopping strength. The parameter $\alpha_{\pm}=\alpha \pm \alpha\,\Gamma/t$, where $\alpha$ and $\alpha\,\Gamma/t$ represent the spin-orbit strength and its non-reciprocity, respectively. As in the HNK model, the non-Hermitian Rashba superconductor  [Eq.~\eqref{eq:SpinfulH}] hosts EPs at $(|\Gamma|,B)=(t,0)$. Moreover, the PHS operator in Eq.~\eqref{eq:MP-site} is given by $\mathcal C = \tau^x \otimes \sigma^0 $. In what follows, we address two case, namely, a uniform non-Hermitian Rashba superconductor and a junction based on such a model but partially left in the normal state.

\subsection{Uniform superconducting single chain }
When superconductivity is homogeneous [Fig.~\ref{fig:Schematics}(c)], the Hermitian regime ($\Gamma=0$) undergoes a topological phase transition at $B_{c}^{\rm H} = \sqrt{\mu^2 + \Delta^2}$~\cite{PhysRevLett.103.020401,PhysRevB.82.134521,PhysRevLett.105.177002,PhysRevLett.105.077001,sato2016majorana,tanaka2024theory} into a topological phase with a pair of MZMs at the system ends. In the non-Hermitian regime ($\Gamma\neq0$) within the thermodynamic limit, the topological phase transition is modified and given by
\begin{equation}\label{eq:NH-TPT}
    B^{\rm NH}_{c}=\sqrt{\mu^2 + 4 \paren{\alpha \Gamma/t} ^2 + \Delta^2}\equiv\sqrt{\tilde\mu^2  + \Delta^2},
\end{equation}
where  $\tilde\mu=\sqrt{\mu^2 + 4 \paren{\alpha \Gamma/t} ^2}$ represents an effective chemical potential. Thus, the first effect of non-Hermiticity via the nonreciprocity $\Gamma$ is to renormalize the chemical potential. Since $B_{\rm c}^{\rm NH}\paren{\Gamma \neq 0}>B^{\rm H}$, the NH model requires a larger  Zeeman field to transition to the topological phase. However, in the regime $\alpha<t,\, 0 <|\Gamma| \ll t$, both critical fields are approximately equal to each other, namely, $B_{c}^{\rm NH} \approx B_{c}^{\rm H}$. We will test the impact of non-Hermiticity on the low-energy physics in this regime.

\begin{figure}[!t]
    \centering
        \includegraphics[width=0.48\textwidth]{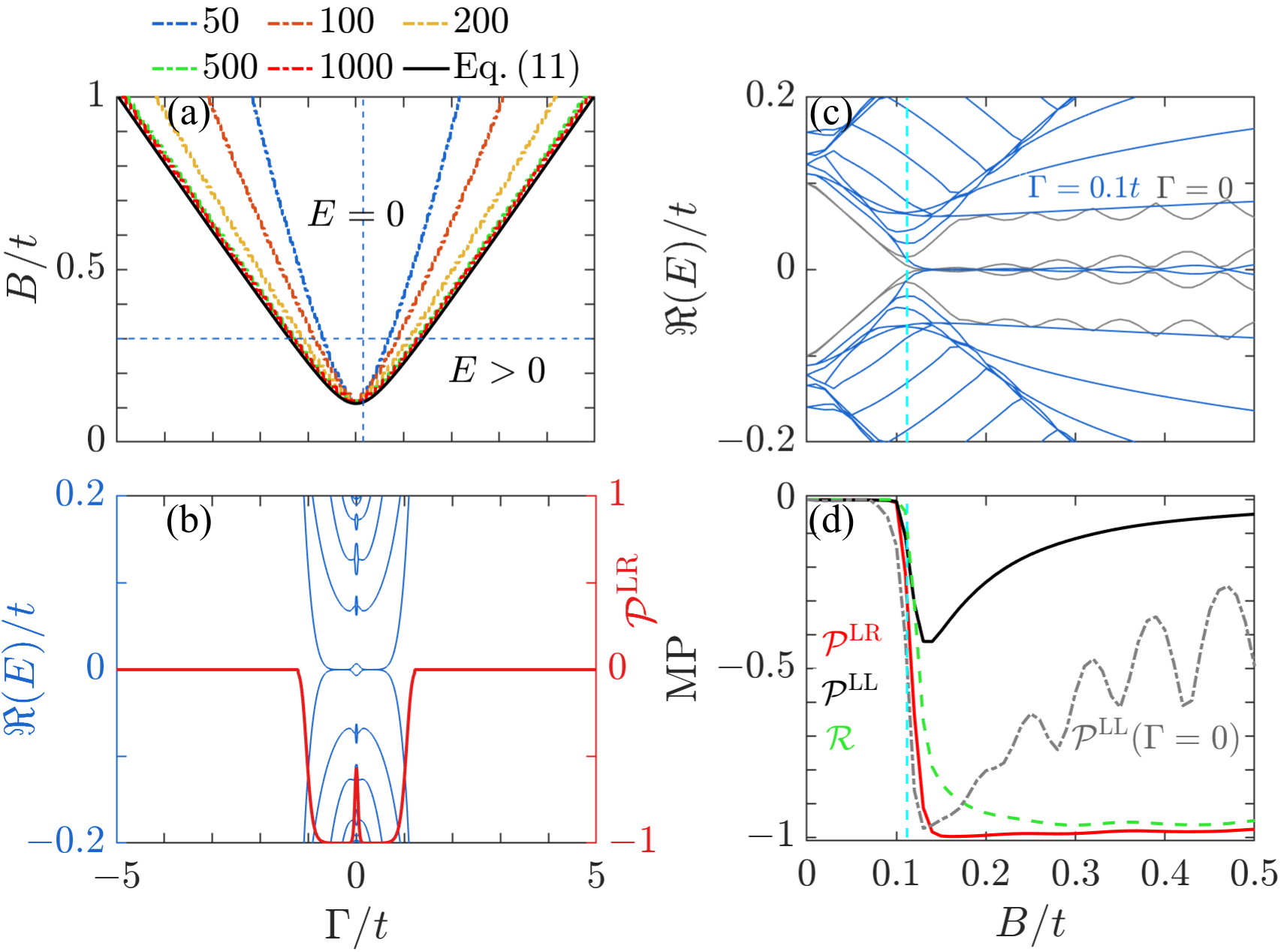}
	\caption{Topology of zero-energy states in the NH Rashba superconducting chain. (a) Size dependence of the phase diagram in $B-\Gamma$ parameter space for several $N_{\rm S}$. The region of $E=0$ is bounded by the curves and approaches Eq.~\eqref{eq:NH-TPT} in the thermodynamic limit. (b) $\Re(E)$ and MP along the dashed horizontal line in (a). (c) $\Re(E)$ along the vertical line in (a) for $\Gamma =0$ (gray) and $\Gamma =0.1t$ (blue). (d) The corresponding MPs of the zero-energy states in (c). Cyan line in (c,d) denote the topological transition point in the Hermitan phase, $B_{\rm c}^{\rm H}$. Here $\Delta = \alpha = 0.1t$  and $\mu = 0.05t$. In (b-c) $N_{\rm S}^{} = 50$.}
	\label{fig:SpinfullChain}
\end{figure}

Having pointed out the role of non-Hermiticity on the topological phase transition, in  Fig.~\ref{fig:SpinfullChain}(a) we visualize this discussion by showing the phase diagram of  zero-energy states in the NH Rashba superconductor. In general, one observes that such a phase diagram exhibits  features similar to those of the Hermitian regime in the $\mu-B$ space, see e.g., Refs.\,\cite{Lutchyn2018Majorana,PhysRevLett.123.117001,PhysRevB.110.165404}. This shows that $\Gamma$ acts like a chemical potential and can be used as an extra knob to adjust the size of the topological phase. The transition point varies with system size and approaches Eq.~\eqref{eq:NH-TPT} in the thermodynamic limit ($N_{\rm S} \ge 500$ for the parameters in  Fig.~\ref{fig:SpinfullChain}).

%
To confirm the topological superconductivity status of the chain, in Fig.~\ref{fig:SpinfullChain}(b) we show the eigenvalues and the nonlocal biorthogonal MP, $\mathcal P^{\rm LR}$, along the horizontal line Fig.~\ref{fig:SpinfullChain}(a). We find that the region enclosed by the curves in Fig.~\ref{fig:SpinfullChain}(a) are topological, since the lowest energy levels are zero and $\PLR=-1$ in this region except at $\Gamma = 0$ where there is a finite energy splitting and $-1<\PLR\ll 0$. The regions outside the bounds ( curves in Fig.~\ref{fig:SpinfullChain}(a)) have trivial topology.  The results also show that, similar to the HNK model of Section~\ref{sec:HNK}, nonreciprocity enhances the MZMs robustness since finite $\Gamma$ drives $\Gamma$ towards $-1$ and lowest energy towards zero within the bounded regions.

To further understand the response of the NH regime of this model, we show in Fig.~\ref{fig:SpinfullChain}(c,d) the evolution of the energy levels and the nonlocal MP indicators as a function of the Zeeman field. The first effect of nonreciprocity ($\Gamma \neq 0$) on the energy levels is that it reduces the energy splitting of Majorana states appearing after the topological phase transition (cyan dashed line), compare blue and gray curves in Fig.~\ref{fig:SpinfullChain}(c). Notably, the nonlocal MP indicators shown in Fig.~\ref{fig:SpinfullChain}(d) directly characterize the topological phase transition and the quality of Majorana states but with important differences and similarities between orthogonal ($\PRR,\, \PLL$) and biorthogonal ($\PLR,\, \PRL$) MPs: while all the nonlocal MP indicators sense the topological phase transition, with $\PLR$ and $\mathcal{R}$ reaching $-1$ in the topological phase, $\PLL$ develops a decay as the Zeeman field increases inside the topological transition because it is not a topological indicator in the NH phase. We also note that the $\PLL$ in the Hermitian regime also captures the oscillations of Majorana states with $B$. Since having MP indicators of $-1$ means that Majorana states have zero energy and are well localized, the obtained results  confirm that non reciprocity enhances the MZMs compared to the Hermitian regime. It is worth mentioning that, while we have presented results for $N_{\rm S}^{}=50$, we find  the same results for longer chains. Next, we study superconductor-normal metal junctions based on the non-Hermitian Rashba superconductor discussed here.

%
\subsection{Non-Hermitian Rashba superconducting junctions}
\begin{figure}[!t]
    \centering
        \includegraphics[width=0.48\textwidth]{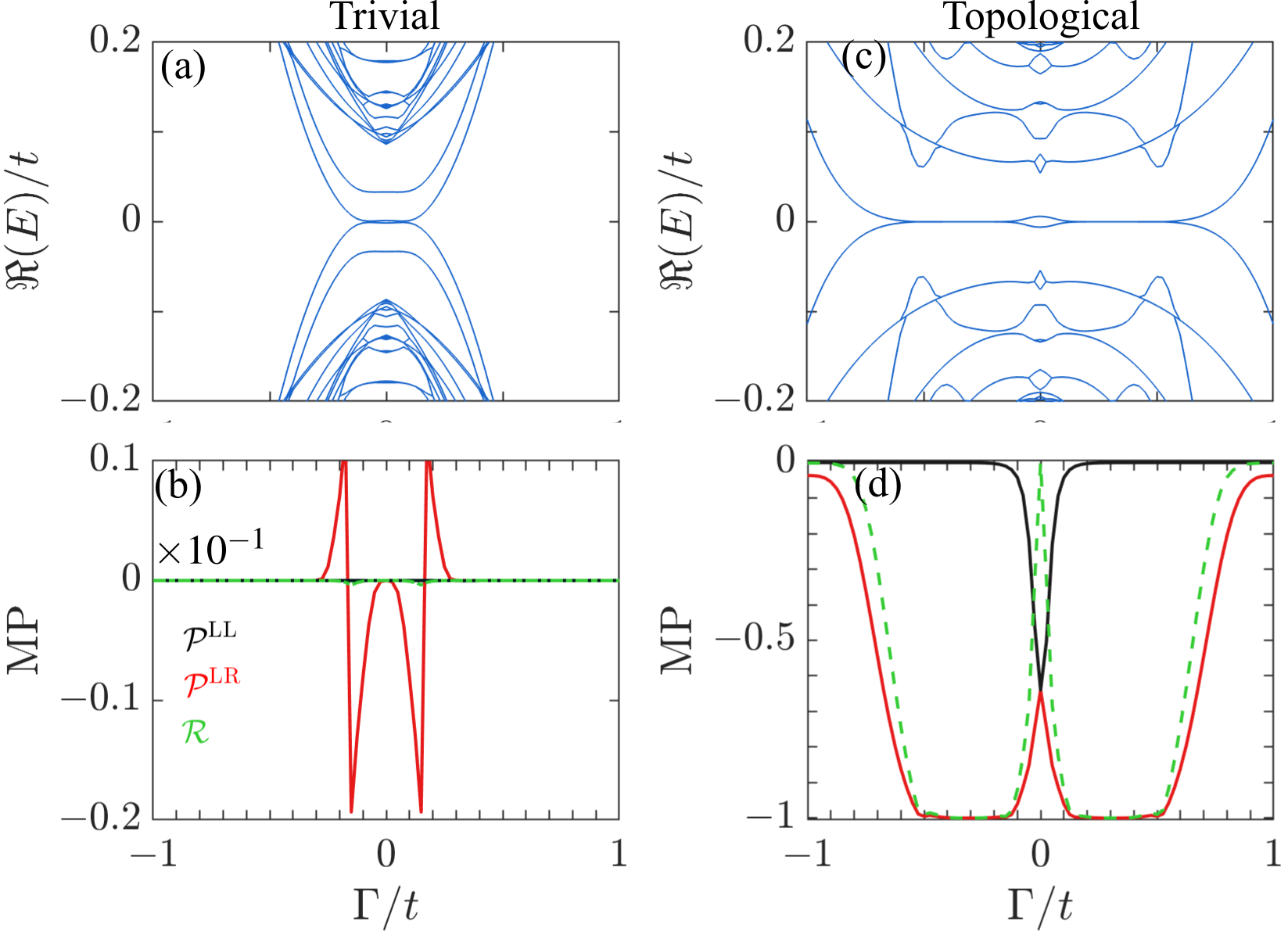}
	\caption{Trivial and nontrivial zero-energy states in SN junctions of NH Rashba SC. (a) $\Re(E)$ as a function of $\Gamma$ in the trivial phase at $B=0.02t$.  (b) Nonlocal MP indicators of the zero-energy state in (a). (c) Same as (a) but in the topological phase for $B =0.3t$. (d) MP components of the zero-energy state in (c). Here $N_{\rm S} = 50,\, N_{\rm n} = 10$  while all other parameters are the same as in Fig.~\ref{fig:SpinfullChain}.
    }
	\label{fig:Spinfull_NS}
\end{figure}

We now address a superconductor-normal junction based on the non-Hermitian Rashba superconductor given by Eq.\,(\ref{eq:SpinfulH}) under nonreciprocity and depicted in Fig.~\ref{fig:Schematics}(d). In the Hermitian regime ($\Gamma=0$), spatially-confined TZESs mimicking MZMs can be trapped in the normal region of the chain~\cite{PhysRevB.91.024514,PhysRevLett.123.117001,PhysRevB.104.L020501,PhysRevB.105.144509,PhysRevB.107.184519,prada2019andreev,PhysRevB.98.245407}. Here, we investigate the impact of $\Gamma$ on TZESs and MZMs and the utility of the nonlocal MP indicators for their identification. For this purpose, we show the real energies and MP indicators as functions of $\Gamma$ in Fig.~\ref{fig:Spinfull_NS} in the trivial ($B<B_{c}^{\rm NH}$) and topological  ($B>B_{c}^{\rm NH}$) phases. 

In the trivial phase ($B<B_{c}^{\rm NH}$),  zero-energy levels appear for a finite range of $\Gamma$ values, see  Fig.~\ref{fig:Spinfull_NS}(a). Within this range, all nonlocal MP  indicators in Fig.~\ref{fig:Spinfull_NS}(b) become $\mathcal{P}^{ab},\,\mathcal{R} \approx 0$,  indicating that these zero-energy levels are non-Majorana trivial states. In contrast, in the topological phase  ($B>B^{\rm H}$),  the formation of zero-energy levels for $\Gamma\neq0$ is characterized by nonlocal MPs indicators approaching values of $-1$, which signals the emergence of fully nonlocal MZMs, see Figs.~\ref{fig:Spinfull_NS}(c,d). Of particular relevance is that  the nonlocal biorthogonal MP,  $\mathcal{P}^{\rm LR}$, is a good topological indicator for NH topological superconductors. 
%
%
\section{Conclusions}\label{sec:Conclusion}
 In this work, we propose and employ the nonlocal Majorana polarization as a diagnostic tool for distinguishing between Majorana zero modes, exceptional points, and trivial zero-energy states in non-Hermitian topological superconductors. Our results demonstrate that, in both spinless and spinfull models, the  nonlocal biorthogonal Majorana polarization provides a reliable means of identifying the nature of low-energy states in non-Hermtian systems, including isolated superconductors and superconductor–normal-metal junctions. Moreover, we find that the robustness of Majorana states is enhanced in the presence of non-Hermiticity. We also present analytic result for  minimal non-Hermitian Kitaev models and show that non-Hermiticity restores MZMs destroyed by chemical potential. This observation may have direct relevance for ongoing experiments on minimal Kitaev chains, where non-Hermitian effects can naturally arise through coupling to measurement probes.

\begin{acknowledgments}
This work is part of the Finnish Centre of Excellence in Quantum Materials (QMAT). O. A. A. acknowledges support from the Research Council of Finland (Projects No. 354735 and 355056) and through the Finnish Quantum Flagship, project number 359240.  J. C. acknowledges financial support from  the Swedish Research Council (Vetenskapsr{\aa}det Grant No. 2021-04121), and from the Olle Engkvist Foundation (Grant No.  243-1026). Part of this work was carried out during the NORDITA Summer Internship Program (2025 session).
\end{acknowledgments}

%
%
\bibliography{biblio}

\begin{thebibliography}{101}%
\makeatletter
\providecommand \@ifxundefined [1]{%
 \@ifx{#1\undefined}
}%
\providecommand \@ifnum [1]{%
 \ifnum #1\expandafter \@firstoftwo
 \else \expandafter \@secondoftwo
 \fi
}%
\providecommand \@ifx [1]{%
 \ifx #1\expandafter \@firstoftwo
 \else \expandafter \@secondoftwo
 \fi
}%
\providecommand \natexlab [1]{#1}%
\providecommand \enquote  [1]{``#1''}%
\providecommand \bibnamefont  [1]{#1}%
\providecommand \bibfnamefont [1]{#1}%
\providecommand \citenamefont [1]{#1}%
\providecommand \href@noop [0]{\@secondoftwo}%
\providecommand \href [0]{\begingroup \@sanitize@url \@href}%
\providecommand \@href[1]{\@@startlink{#1}\@@href}%
\providecommand \@@href[1]{\endgroup#1\@@endlink}%
\providecommand \@sanitize@url [0]{\catcode `\\12\catcode `\$12\catcode
  `\&12\catcode `\#12\catcode `\^12\catcode `\_12\catcode `\%12\relax}%
\providecommand \@@startlink[1]{}%
\providecommand \@@endlink[0]{}%
\providecommand \url  [0]{\begingroup\@sanitize@url \@url }%
\providecommand \@url [1]{\endgroup\@href {#1}{\urlprefix }}%
\providecommand \urlprefix  [0]{URL }%
\providecommand \Eprint [0]{\href }%
\providecommand \doibase [0]{https://doi.org/}%
\providecommand \selectlanguage [0]{\@gobble}%
\providecommand \bibinfo  [0]{\@secondoftwo}%
\providecommand \bibfield  [0]{\@secondoftwo}%
\providecommand \translation [1]{[#1]}%
\providecommand \BibitemOpen [0]{}%
\providecommand \bibitemStop [0]{}%
\providecommand \bibitemNoStop [0]{.\EOS\space}%
\providecommand \EOS [0]{\spacefactor3000\relax}%
\providecommand \BibitemShut  [1]{\csname bibitem#1\endcsname}%
\let\auto@bib@innerbib\@empty
\bibitem [{\citenamefont {Sato}\ and\ \citenamefont
  {Fujimoto}(2016)}]{sato2016majorana}%
  \BibitemOpen
  \bibfield  {author} {\bibinfo {author} {\bibfnamefont {M.}~\bibnamefont
  {Sato}}\ and\ \bibinfo {author} {\bibfnamefont {S.}~\bibnamefont
  {Fujimoto}},\ }\bibfield  {title} {\bibinfo {title} {Majorana fermions and
  topology in superconductors},\ }\href
  {https://doi.org/10.7566/JPSJ.85.072001} {\bibfield  {journal} {\bibinfo
  {journal} {J. Phys. Soc. Jpn.}\ }\textbf {\bibinfo {volume} {85}},\ \bibinfo
  {pages} {072001} (\bibinfo {year} {2016})}\BibitemShut {NoStop}%
\bibitem [{\citenamefont {Aguado}(2017)}]{Aguadoreview17}%
  \BibitemOpen
  \bibfield  {author} {\bibinfo {author} {\bibfnamefont {R.}~\bibnamefont
  {Aguado}},\ }\bibfield  {title} {\bibinfo {title} {Majorana quasiparticles in
  condensed matter},\ }\href {https://doi.org/10.1393/ncr/i2017-10141-9}
  {\bibfield  {journal} {\bibinfo  {journal} {Riv. Nuovo Cimento}\ }\textbf
  {\bibinfo {volume} {40}},\ \bibinfo {pages} {523} (\bibinfo {year}
  {2017})}\BibitemShut {NoStop}%
\bibitem [{\citenamefont {Sato}\ and\ \citenamefont
  {Ando}(2017)}]{sato2017topological}%
  \BibitemOpen
  \bibfield  {author} {\bibinfo {author} {\bibfnamefont {M.}~\bibnamefont
  {Sato}}\ and\ \bibinfo {author} {\bibfnamefont {Y.}~\bibnamefont {Ando}},\
  }\bibfield  {title} {\bibinfo {title} {Topological superconductors: {A}
  review},\ }\href {https://doi.org/10.1088/1361-6633/aa6ac7} {\bibfield
  {journal} {\bibinfo  {journal} {Rep. Prog. Phys.}\ }\textbf {\bibinfo
  {volume} {80}},\ \bibinfo {pages} {076501} (\bibinfo {year}
  {2017})}\BibitemShut {NoStop}%
\bibitem [{\citenamefont {Flensberg}\ \emph {et~al.}(2021)\citenamefont
  {Flensberg}, \citenamefont {von Oppen},\ and\ \citenamefont
  {Stern}}]{flensberg2021engineered}%
  \BibitemOpen
  \bibfield  {author} {\bibinfo {author} {\bibfnamefont {K.}~\bibnamefont
  {Flensberg}}, \bibinfo {author} {\bibfnamefont {F.}~\bibnamefont {von
  Oppen}},\ and\ \bibinfo {author} {\bibfnamefont {A.}~\bibnamefont {Stern}},\
  }\bibfield  {title} {\bibinfo {title} {Engineered platforms for topological
  superconductivity and {M}ajorana zero modes},\ }\href
  {https://www.nature.com/articles/s41578-021-00336-6} {\bibfield  {journal}
  {\bibinfo  {journal} {Nat. Rev. Mater.}\ }\textbf {\bibinfo {volume} {6}},\
  \bibinfo {pages} {944} (\bibinfo {year} {2021})}\BibitemShut {NoStop}%
\bibitem [{\citenamefont {Marra}(2022)}]{Marra_2022}%
  \BibitemOpen
  \bibfield  {author} {\bibinfo {author} {\bibfnamefont {P.}~\bibnamefont
  {Marra}},\ }\bibfield  {title} {\bibinfo {title} {Majorana nanowires for
  topological quantum computation},\ }\href {https://doi.org/10.1063/5.0102999}
  {\bibfield  {journal} {\bibinfo  {journal} {J. Appl. Phys.}\ }\textbf
  {\bibinfo {volume} {132}},\ \bibinfo {pages} {231101} (\bibinfo {year}
  {2022})}\BibitemShut {NoStop}%
\bibitem [{\citenamefont {Tanaka}\ \emph {et~al.}(2024)\citenamefont {Tanaka},
  \citenamefont {Tamura},\ and\ \citenamefont {Cayao}}]{tanaka2024theory}%
  \BibitemOpen
  \bibfield  {author} {\bibinfo {author} {\bibfnamefont {Y.}~\bibnamefont
  {Tanaka}}, \bibinfo {author} {\bibfnamefont {S.}~\bibnamefont {Tamura}},\
  and\ \bibinfo {author} {\bibfnamefont {J.}~\bibnamefont {Cayao}},\ }\bibfield
   {title} {\bibinfo {title} {Theory of {M}ajorana zero modes in unconventional
  superconductors},\ }\href {https://doi.org/10.1093/ptep/ptae065} {\bibfield
  {journal} {\bibinfo  {journal} {Prog. Theor. Exp. Phys.}\ }\textbf {\bibinfo
  {volume} {2024}},\ \bibinfo {pages} {08C105} (\bibinfo {year}
  {2024})}\BibitemShut {NoStop}%
\bibitem [{\citenamefont {Fukaya}\ \emph {et~al.}(2025)\citenamefont {Fukaya},
  \citenamefont {Lu}, \citenamefont {Yada}, \citenamefont {Tanaka},\ and\
  \citenamefont {Cayao}}]{FukayaJPCM2025}%
  \BibitemOpen
  \bibfield  {author} {\bibinfo {author} {\bibfnamefont {Y.}~\bibnamefont
  {Fukaya}}, \bibinfo {author} {\bibfnamefont {B.}~\bibnamefont {Lu}}, \bibinfo
  {author} {\bibfnamefont {K.}~\bibnamefont {Yada}}, \bibinfo {author}
  {\bibfnamefont {Y.}~\bibnamefont {Tanaka}},\ and\ \bibinfo {author}
  {\bibfnamefont {J.}~\bibnamefont {Cayao}},\ }\bibfield  {title} {\bibinfo
  {title} {Superconducting phenomena in systems with unconventional magnets},\
  }\href {https://iopscience.iop.org/article/10.1088/1361-648X/adf1cf}
  {\bibfield  {journal} {\bibinfo  {journal} {J. Phys.: Condens. Matter}\
  }\textbf {\bibinfo {volume} {37}},\ \bibinfo {pages} {313003} (\bibinfo
  {year} {2025})}\BibitemShut {NoStop}%
\bibitem [{\citenamefont {Kitaev}(2001)}]{kitaev2001unpaired}%
  \BibitemOpen
  \bibfield  {author} {\bibinfo {author} {\bibfnamefont {A.~Y.}\ \bibnamefont
  {Kitaev}},\ }\bibfield  {title} {\bibinfo {title} {Unpaired {Majorana}
  fermions in quantum wires},\ }\href
  {https://doi.org/10.1070/1063-7869/44/10S/S29} {\bibfield  {journal}
  {\bibinfo  {journal} {Phys.-Usp.}\ }\textbf {\bibinfo {volume} {44}},\
  \bibinfo {pages} {131} (\bibinfo {year} {2001})}\BibitemShut {NoStop}%
\bibitem [{\citenamefont {Awoga}\ and\ \citenamefont
  {Cayao}(2024)}]{PhysRevB.110.165404}%
  \BibitemOpen
  \bibfield  {author} {\bibinfo {author} {\bibfnamefont {O.~A.}\ \bibnamefont
  {Awoga}}\ and\ \bibinfo {author} {\bibfnamefont {J.}~\bibnamefont {Cayao}},\
  }\bibfield  {title} {\bibinfo {title} {Identifying trivial and {M}ajorana
  zero-energy modes using the {M}ajorana polarization},\ }\href
  {https://doi.org/10.1103/PhysRevB.110.165404} {\bibfield  {journal} {\bibinfo
   {journal} {Phys. Rev. B}\ }\textbf {\bibinfo {volume} {110}},\ \bibinfo
  {pages} {165404} (\bibinfo {year} {2024})}\BibitemShut {NoStop}%
\bibitem [{\citenamefont {Ashida}\ \emph {et~al.}(2020)\citenamefont {Ashida},
  \citenamefont {Gong},\ and\ \citenamefont
  {Ueda}}]{doi:10.1080/00018732.2021.1876991}%
  \BibitemOpen
  \bibfield  {author} {\bibinfo {author} {\bibfnamefont {Y.}~\bibnamefont
  {Ashida}}, \bibinfo {author} {\bibfnamefont {Z.}~\bibnamefont {Gong}},\ and\
  \bibinfo {author} {\bibfnamefont {M.}~\bibnamefont {Ueda}},\ }\bibfield
  {title} {\bibinfo {title} {Non-{H}ermitian physics},\ }\href
  {https://doi.org/10.1080/00018732.2021.1876991} {\bibfield  {journal}
  {\bibinfo  {journal} {Adv. Phys.}\ }\textbf {\bibinfo {volume} {69}},\
  \bibinfo {pages} {249} (\bibinfo {year} {2020})}\BibitemShut {NoStop}%
\bibitem [{\citenamefont {Okuma}\ and\ \citenamefont {Sato}(2023)}]{Okuma2023}%
  \BibitemOpen
  \bibfield  {author} {\bibinfo {author} {\bibfnamefont {N.}~\bibnamefont
  {Okuma}}\ and\ \bibinfo {author} {\bibfnamefont {M.}~\bibnamefont {Sato}},\
  }\bibfield  {title} {\bibinfo {title} {Non-{H}ermitian topological phenomena:
  {A} review},\ }\href
  {https://doi.org/10.1146/annurev-conmatphys-040521-033133} {\bibfield
  {journal} {\bibinfo  {journal} {Annu. Rev. Condens. Matter Phys.}\ ,\
  \bibinfo {pages} {83}} (\bibinfo {year} {2023})}\BibitemShut {NoStop}%
\bibitem [{\citenamefont {Bessho}\ \emph {et~al.}(2019)\citenamefont {Bessho},
  \citenamefont {Kawabata},\ and\ \citenamefont
  {Sato}}]{doi:10.7566/JPSCP.30.011098}%
  \BibitemOpen
  \bibfield  {author} {\bibinfo {author} {\bibfnamefont {T.}~\bibnamefont
  {Bessho}}, \bibinfo {author} {\bibfnamefont {K.}~\bibnamefont {Kawabata}},\
  and\ \bibinfo {author} {\bibfnamefont {M.}~\bibnamefont {Sato}},\ }\bibinfo
  {title} {Topological classificaton of non-{H}ermitian gapless phases:
  Exceptional points and bulk {F}ermi arcs},\ in\ \href
  {https://doi.org/10.7566/JPSCP.30.011098} {\emph {\bibinfo {booktitle} {Proc.
  Int. Conf. on Strongly Correlated Electron Systems (SCES2019)}}}\ (\bibinfo
  {publisher} {Physical Society of Japan},\ \bibinfo {year} {2019})\
  Chap.~\bibinfo {chapter} {30}, p.\ \bibinfo {pages} {011098}\BibitemShut
  {NoStop}%
\bibitem [{\citenamefont {Bergholtz}\ \emph {et~al.}(2021)\citenamefont
  {Bergholtz}, \citenamefont {Budich},\ and\ \citenamefont
  {Kunst}}]{RevModPhys.93.015005}%
  \BibitemOpen
  \bibfield  {author} {\bibinfo {author} {\bibfnamefont {E.~J.}\ \bibnamefont
  {Bergholtz}}, \bibinfo {author} {\bibfnamefont {J.~C.}\ \bibnamefont
  {Budich}},\ and\ \bibinfo {author} {\bibfnamefont {F.~K.}\ \bibnamefont
  {Kunst}},\ }\bibfield  {title} {\bibinfo {title} {Exceptional topology of
  non-{H}ermitian systems},\ }\href
  {https://doi.org/10.1103/RevModPhys.93.015005} {\bibfield  {journal}
  {\bibinfo  {journal} {Rev. Mod. Phys.}\ }\textbf {\bibinfo {volume} {93}},\
  \bibinfo {pages} {015005} (\bibinfo {year} {2021})}\BibitemShut {NoStop}%
\bibitem [{\citenamefont {San-Jos\'{e}}\ \emph {et~al.}(2016)\citenamefont
  {San-Jos\'{e}}, \citenamefont {Cayao}, \citenamefont {Prada},\ and\
  \citenamefont {Aguado}}]{JorgeEPs}%
  \BibitemOpen
  \bibfield  {author} {\bibinfo {author} {\bibfnamefont {P.}~\bibnamefont
  {San-Jos\'{e}}}, \bibinfo {author} {\bibfnamefont {J.}~\bibnamefont {Cayao}},
  \bibinfo {author} {\bibfnamefont {E.}~\bibnamefont {Prada}},\ and\ \bibinfo
  {author} {\bibfnamefont {R.}~\bibnamefont {Aguado}},\ }\bibfield  {title}
  {\bibinfo {title} {Majorana bound states from exceptional points in
  non-topological superconductors},\ }\href
  {http://dx.doi.org/10.1038/srep21427} {\bibfield  {journal} {\bibinfo
  {journal} {Sci. Rep.}\ }\textbf {\bibinfo {volume} {6}},\ \bibinfo {pages}
  {21427} (\bibinfo {year} {2016})}\BibitemShut {NoStop}%
\bibitem [{\citenamefont {Cayao}\ and\ \citenamefont
  {Black-Schaffer}(2022)}]{PhysRevB.105.094502}%
  \BibitemOpen
  \bibfield  {author} {\bibinfo {author} {\bibfnamefont {J.}~\bibnamefont
  {Cayao}}\ and\ \bibinfo {author} {\bibfnamefont {A.~M.}\ \bibnamefont
  {Black-Schaffer}},\ }\bibfield  {title} {\bibinfo {title} {Exceptional
  odd-frequency pairing in non-{H}ermitian superconducting systems},\ }\href
  {https://doi.org/10.1103/PhysRevB.105.094502} {\bibfield  {journal} {\bibinfo
   {journal} {Phys. Rev. B}\ }\textbf {\bibinfo {volume} {105}},\ \bibinfo
  {pages} {094502} (\bibinfo {year} {2022})}\BibitemShut {NoStop}%
\bibitem [{\citenamefont {Cayao}\ and\ \citenamefont
  {Black-Schaffer}(2023)}]{PhysRevB.107.104515}%
  \BibitemOpen
  \bibfield  {author} {\bibinfo {author} {\bibfnamefont {J.}~\bibnamefont
  {Cayao}}\ and\ \bibinfo {author} {\bibfnamefont {A.~M.}\ \bibnamefont
  {Black-Schaffer}},\ }\bibfield  {title} {\bibinfo {title} {Bulk {B}ogoliubov
  {F}ermi arcs in non-{H}ermitian superconducting systems},\ }\href
  {https://doi.org/10.1103/PhysRevB.107.104515} {\bibfield  {journal} {\bibinfo
   {journal} {Phys. Rev. B}\ }\textbf {\bibinfo {volume} {107}},\ \bibinfo
  {pages} {104515} (\bibinfo {year} {2023})}\BibitemShut {NoStop}%
\bibitem [{\citenamefont {Zyuzin}\ and\ \citenamefont
  {Simon}(2019)}]{PhysRevB.99.165145}%
  \BibitemOpen
  \bibfield  {author} {\bibinfo {author} {\bibfnamefont {A.~A.}\ \bibnamefont
  {Zyuzin}}\ and\ \bibinfo {author} {\bibfnamefont {P.}~\bibnamefont {Simon}},\
  }\bibfield  {title} {\bibinfo {title} {Disorder-induced exceptional points
  and nodal lines in {D}irac superconductors},\ }\href
  {https://doi.org/10.1103/PhysRevB.99.165145} {\bibfield  {journal} {\bibinfo
  {journal} {Phys. Rev. B}\ }\textbf {\bibinfo {volume} {99}},\ \bibinfo
  {pages} {165145} (\bibinfo {year} {2019})}\BibitemShut {NoStop}%
\bibitem [{\citenamefont {Okuma}\ and\ \citenamefont
  {Sato}(2019)}]{PhysRevLett.123.097701}%
  \BibitemOpen
  \bibfield  {author} {\bibinfo {author} {\bibfnamefont {N.}~\bibnamefont
  {Okuma}}\ and\ \bibinfo {author} {\bibfnamefont {M.}~\bibnamefont {Sato}},\
  }\bibfield  {title} {\bibinfo {title} {Topological phase transition driven by
  infinitesimal instability: {M}ajorana fermions in non-{H}ermitian
  spintronics},\ }\href {https://doi.org/10.1103/PhysRevLett.123.097701}
  {\bibfield  {journal} {\bibinfo  {journal} {Phys. Rev. Lett.}\ }\textbf
  {\bibinfo {volume} {123}},\ \bibinfo {pages} {097701} (\bibinfo {year}
  {2019})}\BibitemShut {NoStop}%
\bibitem [{\citenamefont {Jana}\ \emph {et~al.}(2021)\citenamefont {Jana},
  \citenamefont {Chowdhury},\ and\ \citenamefont {Saha}}]{PhysRevB.103.235438}%
  \BibitemOpen
  \bibfield  {author} {\bibinfo {author} {\bibfnamefont {S.}~\bibnamefont
  {Jana}}, \bibinfo {author} {\bibfnamefont {D.}~\bibnamefont {Chowdhury}},\
  and\ \bibinfo {author} {\bibfnamefont {A.}~\bibnamefont {Saha}},\ }\bibfield
  {title} {\bibinfo {title} {Emergence of exceptional points and their
  spectroscopic signature in a {D}irac semimetal--dirty superconductor
  heterojunction},\ }\href {https://doi.org/10.1103/PhysRevB.103.235438}
  {\bibfield  {journal} {\bibinfo  {journal} {Phys. Rev. B}\ }\textbf {\bibinfo
  {volume} {103}},\ \bibinfo {pages} {235438} (\bibinfo {year}
  {2021})}\BibitemShut {NoStop}%
\bibitem [{\citenamefont {Kornich}\ and\ \citenamefont
  {Trauzettel}(2022)}]{PhysRevResearch.4.L022018}%
  \BibitemOpen
  \bibfield  {author} {\bibinfo {author} {\bibfnamefont {V.}~\bibnamefont
  {Kornich}}\ and\ \bibinfo {author} {\bibfnamefont {B.}~\bibnamefont
  {Trauzettel}},\ }\bibfield  {title} {\bibinfo {title} {Signature of
  $\mathcal{P}\mathcal{T}$-symmetric non-{H}ermitian superconductivity in
  angle-resolved photoelectron fluctuation spectroscopy},\ }\href
  {https://doi.org/10.1103/PhysRevResearch.4.L022018} {\bibfield  {journal}
  {\bibinfo  {journal} {Phys. Rev. Research}\ }\textbf {\bibinfo {volume}
  {4}},\ \bibinfo {pages} {L022018} (\bibinfo {year} {2022})}\BibitemShut
  {NoStop}%
\bibitem [{\citenamefont {Cayao}(2024{\natexlab{a}})}]{PhysRevB.110.085414}%
  \BibitemOpen
  \bibfield  {author} {\bibinfo {author} {\bibfnamefont {J.}~\bibnamefont
  {Cayao}},\ }\bibfield  {title} {\bibinfo {title} {Non-{H}ermitian zero-energy
  pinning of {A}ndreev and {M}ajorana bound states in
  superconductor-semiconductor systems},\ }\href
  {https://doi.org/10.1103/PhysRevB.110.085414} {\bibfield  {journal} {\bibinfo
   {journal} {Phys. Rev. B}\ }\textbf {\bibinfo {volume} {110}},\ \bibinfo
  {pages} {085414} (\bibinfo {year} {2024}{\natexlab{a}})}\BibitemShut
  {NoStop}%
\bibitem [{\citenamefont {Cayao}\ and\ \citenamefont
  {Sato}(2024{\natexlab{a}})}]{cayao2023nonhermitian}%
  \BibitemOpen
  \bibfield  {author} {\bibinfo {author} {\bibfnamefont {J.}~\bibnamefont
  {Cayao}}\ and\ \bibinfo {author} {\bibfnamefont {M.}~\bibnamefont {Sato}},\
  }\bibfield  {title} {\bibinfo {title} {Non-{H}ermitian phase-biased
  {J}osephson junctions},\ }\href
  {https://doi.org/10.1103/PhysRevB.110.L201403} {\bibfield  {journal}
  {\bibinfo  {journal} {Phys. Rev. B}\ }\textbf {\bibinfo {volume} {110}},\
  \bibinfo {pages} {L201403} (\bibinfo {year}
  {2024}{\natexlab{a}})}\BibitemShut {NoStop}%
\bibitem [{\citenamefont {Li}\ \emph {et~al.}(2024)\citenamefont {Li},
  \citenamefont {Sun},\ and\ \citenamefont {Trauzettel}}]{li2023anomalous}%
  \BibitemOpen
  \bibfield  {author} {\bibinfo {author} {\bibfnamefont {C.-A.}\ \bibnamefont
  {Li}}, \bibinfo {author} {\bibfnamefont {H.-P.}\ \bibnamefont {Sun}},\ and\
  \bibinfo {author} {\bibfnamefont {B.}~\bibnamefont {Trauzettel}},\ }\bibfield
   {title} {\bibinfo {title} {Anomalous {A}ndreev spectrum and transport in
  non-{H}ermitian {J}osephson junctions},\ }\href
  {https://doi.org/10.1103/PhysRevB.109.214514} {\bibfield  {journal} {\bibinfo
   {journal} {Phys. Rev. B}\ }\textbf {\bibinfo {volume} {109}},\ \bibinfo
  {pages} {214514} (\bibinfo {year} {2024})}\BibitemShut {NoStop}%
\bibitem [{\citenamefont {Avila}\ \emph {et~al.}(2019)\citenamefont {Avila},
  \citenamefont {Pe{\~n}aranda}, \citenamefont {Prada}, \citenamefont
  {San-Jose},\ and\ \citenamefont {Aguado}}]{avila2019non}%
  \BibitemOpen
  \bibfield  {author} {\bibinfo {author} {\bibfnamefont {J.}~\bibnamefont
  {Avila}}, \bibinfo {author} {\bibfnamefont {F.}~\bibnamefont
  {Pe{\~n}aranda}}, \bibinfo {author} {\bibfnamefont {E.}~\bibnamefont
  {Prada}}, \bibinfo {author} {\bibfnamefont {P.}~\bibnamefont {San-Jose}},\
  and\ \bibinfo {author} {\bibfnamefont {R.}~\bibnamefont {Aguado}},\
  }\bibfield  {title} {\bibinfo {title} {Non-{H}ermitian topology as a unifying
  framework for the {A}ndreev versus {M}ajorana states controversy},\ }\href
  {https://www.nature.com/articles/s42005-019-0231-8} {\bibfield  {journal}
  {\bibinfo  {journal} {Commun. Phys.}\ }\textbf {\bibinfo {volume} {2}},\
  \bibinfo {pages} {133} (\bibinfo {year} {2019})}\BibitemShut {NoStop}%
\bibitem [{\citenamefont {Arouca}\ \emph {et~al.}(2023)\citenamefont {Arouca},
  \citenamefont {Cayao},\ and\ \citenamefont
  {Black-Schaffer}}]{PhysRevB.108.L060506}%
  \BibitemOpen
  \bibfield  {author} {\bibinfo {author} {\bibfnamefont {R.}~\bibnamefont
  {Arouca}}, \bibinfo {author} {\bibfnamefont {J.}~\bibnamefont {Cayao}},\ and\
  \bibinfo {author} {\bibfnamefont {A.~M.}\ \bibnamefont {Black-Schaffer}},\
  }\bibfield  {title} {\bibinfo {title} {Topological superconductivity enhanced
  by exceptional points},\ }\href
  {https://doi.org/10.1103/PhysRevB.108.L060506} {\bibfield  {journal}
  {\bibinfo  {journal} {Phys. Rev. B}\ }\textbf {\bibinfo {volume} {108}},\
  \bibinfo {pages} {L060506} (\bibinfo {year} {2023})}\BibitemShut {NoStop}%
\bibitem [{\citenamefont {Shen}\ \emph {et~al.}(2024)\citenamefont {Shen},
  \citenamefont {Lu}, \citenamefont {Lado},\ and\ \citenamefont
  {Trif}}]{shen2024nonhermitian}%
  \BibitemOpen
  \bibfield  {author} {\bibinfo {author} {\bibfnamefont {P.-X.}\ \bibnamefont
  {Shen}}, \bibinfo {author} {\bibfnamefont {Z.}~\bibnamefont {Lu}}, \bibinfo
  {author} {\bibfnamefont {J.~L.}\ \bibnamefont {Lado}},\ and\ \bibinfo
  {author} {\bibfnamefont {M.}~\bibnamefont {Trif}},\ }\bibfield  {title}
  {\bibinfo {title} {Non-{H}ermitian {F}ermi-{D}irac distribution in persistent
  current transport},\ }\href {https://doi.org/10.1103/PhysRevLett.133.086301}
  {\bibfield  {journal} {\bibinfo  {journal} {Phys. Rev. Lett.}\ }\textbf
  {\bibinfo {volume} {133}},\ \bibinfo {pages} {086301} (\bibinfo {year}
  {2024})}\BibitemShut {NoStop}%
\bibitem [{\citenamefont {Pino}\ \emph {et~al.}(2025)\citenamefont {Pino},
  \citenamefont {Meir},\ and\ \citenamefont {Aguado}}]{pino2024thermodynamics}%
  \BibitemOpen
  \bibfield  {author} {\bibinfo {author} {\bibfnamefont {D.~M.}\ \bibnamefont
  {Pino}}, \bibinfo {author} {\bibfnamefont {Y.}~\bibnamefont {Meir}},\ and\
  \bibinfo {author} {\bibfnamefont {R.}~\bibnamefont {Aguado}},\ }\bibfield
  {title} {\bibinfo {title} {Thermodynamics of non-{H}ermitian {J}osephson
  junctions with exceptional points},\ }\href
  {https://doi.org/10.1103/PhysRevB.111.L140503} {\bibfield  {journal}
  {\bibinfo  {journal} {Phys. Rev. B}\ }\textbf {\bibinfo {volume} {111}},\
  \bibinfo {pages} {L140503} (\bibinfo {year} {2025})}\BibitemShut {NoStop}%
\bibitem [{\citenamefont {Ohnmacht}\ \emph {et~al.}(2025)\citenamefont
  {Ohnmacht}, \citenamefont {Wilhelm}, \citenamefont {Weisbrich},\ and\
  \citenamefont {Belzig}}]{Ohnmacht2024}%
  \BibitemOpen
  \bibfield  {author} {\bibinfo {author} {\bibfnamefont {D.~C.}\ \bibnamefont
  {Ohnmacht}}, \bibinfo {author} {\bibfnamefont {V.}~\bibnamefont {Wilhelm}},
  \bibinfo {author} {\bibfnamefont {H.}~\bibnamefont {Weisbrich}},\ and\
  \bibinfo {author} {\bibfnamefont {W.}~\bibnamefont {Belzig}},\ }\bibfield
  {title} {\bibinfo {title} {Non-{H}ermitian topology in multiterminal
  superconducting junctions},\ }\href
  {https://doi.org/10.1103/PhysRevLett.134.156601} {\bibfield  {journal}
  {\bibinfo  {journal} {Phys. Rev. Lett.}\ }\textbf {\bibinfo {volume} {134}},\
  \bibinfo {pages} {156601} (\bibinfo {year} {2025})}\BibitemShut {NoStop}%
\bibitem [{\citenamefont {Cayao}\ and\ \citenamefont
  {Sato}(2024{\natexlab{b}})}]{cayao2024nonMulti}%
  \BibitemOpen
  \bibfield  {author} {\bibinfo {author} {\bibfnamefont {J.}~\bibnamefont
  {Cayao}}\ and\ \bibinfo {author} {\bibfnamefont {M.}~\bibnamefont {Sato}},\
  }\bibfield  {title} {\bibinfo {title} {Non-{H}ermitian multiterminal
  phase-biased {J}osephson junctions},\ }\href
  {https://doi.org/10.1103/PhysRevB.110.235426} {\bibfield  {journal} {\bibinfo
   {journal} {Phys. Rev. B}\ }\textbf {\bibinfo {volume} {110}},\ \bibinfo
  {pages} {235426} (\bibinfo {year} {2024}{\natexlab{b}})}\BibitemShut
  {NoStop}%
\bibitem [{\citenamefont {Li}\ and\ \citenamefont
  {Trauzettel}(2025)}]{li2025EP}%
  \BibitemOpen
  \bibfield  {author} {\bibinfo {author} {\bibfnamefont {C.-A.}\ \bibnamefont
  {Li}}\ and\ \bibinfo {author} {\bibfnamefont {B.}~\bibnamefont
  {Trauzettel}},\ }\bibfield  {title} {\bibinfo {title} {Exceptional {A}ndreev
  spectrum and supercurrent in $p$-wave non-{H}ermitian {J}osephson
  junctions},\ }\href {https://doi.org/10.1103/58vd-b181} {\bibfield  {journal}
  {\bibinfo  {journal} {Phys. Rev. B}\ }\textbf {\bibinfo {volume} {112}},\
  \bibinfo {pages} {184504} (\bibinfo {year} {2025})}\BibitemShut {NoStop}%
\bibitem [{\citenamefont {Capecelatro}\ \emph {et~al.}(2025)\citenamefont
  {Capecelatro}, \citenamefont {Marciani}, \citenamefont {Campagnano},\ and\
  \citenamefont {Lucignano}}]{PhysRevB.111.064517}%
  \BibitemOpen
  \bibfield  {author} {\bibinfo {author} {\bibfnamefont {R.}~\bibnamefont
  {Capecelatro}}, \bibinfo {author} {\bibfnamefont {M.}~\bibnamefont
  {Marciani}}, \bibinfo {author} {\bibfnamefont {G.}~\bibnamefont
  {Campagnano}},\ and\ \bibinfo {author} {\bibfnamefont {P.}~\bibnamefont
  {Lucignano}},\ }\bibfield  {title} {\bibinfo {title} {Andreev non-{H}ermitian
  {H}amiltonian for open {J}osephson junctions from {G}reen's functions},\
  }\href {https://doi.org/10.1103/PhysRevB.111.064517} {\bibfield  {journal}
  {\bibinfo  {journal} {Phys. Rev. B}\ }\textbf {\bibinfo {volume} {111}},\
  \bibinfo {pages} {064517} (\bibinfo {year} {2025})}\BibitemShut {NoStop}%
\bibitem [{\citenamefont {Ogino}\ and\ \citenamefont
  {Uchino}(2025)}]{ogino2025}%
  \BibitemOpen
  \bibfield  {author} {\bibinfo {author} {\bibfnamefont {R.}~\bibnamefont
  {Ogino}}\ and\ \bibinfo {author} {\bibfnamefont {S.}~\bibnamefont {Uchino}},\
  }\bibfield  {title} {\bibinfo {title} {Anomalous supercurrents in the
  presence of particle losses},\ }\href {https://arxiv.org/abs/2505.21085}
  {\bibfield  {journal} {\bibinfo  {journal} {arXiv:2505.21085}\ } (\bibinfo
  {year} {2025})}\BibitemShut {NoStop}%
\bibitem [{\citenamefont {Solow}\ and\ \citenamefont
  {Flensberg}(2025)}]{solow2025EP}%
  \BibitemOpen
  \bibfield  {author} {\bibinfo {author} {\bibfnamefont {O.}~\bibnamefont
  {Solow}}\ and\ \bibinfo {author} {\bibfnamefont {K.}~\bibnamefont
  {Flensberg}},\ }\bibfield  {title} {\bibinfo {title} {Signatures of
  exceptional points in multiterminal superconductor--normal metal junctions},\
  }\href {https://doi.org/10.1103/dmfm-71l6} {\bibfield  {journal} {\bibinfo
  {journal} {Phys. Rev. B}\ }\textbf {\bibinfo {volume} {112}},\ \bibinfo
  {pages} {L161402} (\bibinfo {year} {2025})}\BibitemShut {NoStop}%
\bibitem [{\citenamefont {Qi}\ \emph {et~al.}(2025)\citenamefont {Qi},
  \citenamefont {Lu}, \citenamefont {Liu}, \citenamefont {Chen},\ and\
  \citenamefont {Xie}}]{JunjieNHDiode}%
  \BibitemOpen
  \bibfield  {author} {\bibinfo {author} {\bibfnamefont {J.}~\bibnamefont
  {Qi}}, \bibinfo {author} {\bibfnamefont {M.}~\bibnamefont {Lu}}, \bibinfo
  {author} {\bibfnamefont {J.}~\bibnamefont {Liu}}, \bibinfo {author}
  {\bibfnamefont {C.-Z.}\ \bibnamefont {Chen}},\ and\ \bibinfo {author}
  {\bibfnamefont {X.~C.}\ \bibnamefont {Xie}},\ }\bibfield  {title} {\bibinfo
  {title} {Non-{H}ermitian superconducting diode effect},\ }\href
  {https://doi.org/10.1103/n51c-17pn} {\bibfield  {journal} {\bibinfo
  {journal} {Phys. Rev. B}\ }\textbf {\bibinfo {volume} {112}},\ \bibinfo
  {pages} {L060502} (\bibinfo {year} {2025})}\BibitemShut {NoStop}%
\bibitem [{\citenamefont {Cayao}\ and\ \citenamefont
  {Sato}(2026{\natexlab{a}})}]{cayaoSatoNH4MZMs}%
  \BibitemOpen
  \bibfield  {author} {\bibinfo {author} {\bibfnamefont {J.}~\bibnamefont
  {Cayao}}\ and\ \bibinfo {author} {\bibfnamefont {M.}~\bibnamefont {Sato}},\
  }\bibfield  {title} {\bibinfo {title} {Non-{H}ermitian {J}osephson junctions
  with four {M}ajorana zero modes},\ }\href
  {https://doi.org/10.7566/JPSJ.95.014705} {\bibfield  {journal} {\bibinfo
  {journal} {J. Phys. Soc. Jpn.}\ }\textbf {\bibinfo {volume} {95}},\ \bibinfo
  {pages} {014705} (\bibinfo {year} {2026}{\natexlab{a}})}\BibitemShut
  {NoStop}%
\bibitem [{\citenamefont {Cayao}\ and\ \citenamefont
  {Aguado}(2025)}]{cayaominimalkitaev}%
  \BibitemOpen
  \bibfield  {author} {\bibinfo {author} {\bibfnamefont {J.}~\bibnamefont
  {Cayao}}\ and\ \bibinfo {author} {\bibfnamefont {R.}~\bibnamefont {Aguado}},\
  }\bibfield  {title} {\bibinfo {title} {Non-{H}ermitian minimal {K}itaev
  chains},\ }\href {https://doi.org/10.1103/PhysRevB.111.205432} {\bibfield
  {journal} {\bibinfo  {journal} {Phys. Rev. B}\ }\textbf {\bibinfo {volume}
  {111}},\ \bibinfo {pages} {205432} (\bibinfo {year} {2025})}\BibitemShut
  {NoStop}%
\bibitem [{\citenamefont {Ezawa}(2024)}]{PhysRevB.109.L161404}%
  \BibitemOpen
  \bibfield  {author} {\bibinfo {author} {\bibfnamefont {M.}~\bibnamefont
  {Ezawa}},\ }\bibfield  {title} {\bibinfo {title} {Even-odd effect on
  robustness of {M}ajorana edge states in short {K}itaev chains},\ }\href
  {https://doi.org/10.1103/PhysRevB.109.L161404} {\bibfield  {journal}
  {\bibinfo  {journal} {Phys. Rev. B}\ }\textbf {\bibinfo {volume} {109}},\
  \bibinfo {pages} {L161404} (\bibinfo {year} {2024})}\BibitemShut {NoStop}%
\bibitem [{\citenamefont {Pay\'a}\ \emph {et~al.}(2026)\citenamefont {Pay\'a},
  \citenamefont {Solow}, \citenamefont {Prada}, \citenamefont {Aguado},\ and\
  \citenamefont {Flensberg}}]{9jdy-b418}%
  \BibitemOpen
  \bibfield  {author} {\bibinfo {author} {\bibfnamefont {C.}~\bibnamefont
  {Pay\'a}}, \bibinfo {author} {\bibfnamefont {O.}~\bibnamefont {Solow}},
  \bibinfo {author} {\bibfnamefont {E.}~\bibnamefont {Prada}}, \bibinfo
  {author} {\bibfnamefont {R.}~\bibnamefont {Aguado}},\ and\ \bibinfo {author}
  {\bibfnamefont {K.}~\bibnamefont {Flensberg}},\ }\bibfield  {title} {\bibinfo
  {title} {Non-{H}ermitian skin effect and electronic nonlocal transport},\
  }\href {https://doi.org/10.1103/9jdy-b418} {\bibfield  {journal} {\bibinfo
  {journal} {Phys. Rev. B}\ }\textbf {\bibinfo {volume} {113}},\ \bibinfo
  {pages} {L161405} (\bibinfo {year} {2026})}\BibitemShut {NoStop}%
\bibitem [{\citenamefont {Kawabata}\ \emph {et~al.}(2019)\citenamefont
  {Kawabata}, \citenamefont {Shiozaki}, \citenamefont {Ueda},\ and\
  \citenamefont {Sato}}]{PhysRevX.9.041015}%
  \BibitemOpen
  \bibfield  {author} {\bibinfo {author} {\bibfnamefont {K.}~\bibnamefont
  {Kawabata}}, \bibinfo {author} {\bibfnamefont {K.}~\bibnamefont {Shiozaki}},
  \bibinfo {author} {\bibfnamefont {M.}~\bibnamefont {Ueda}},\ and\ \bibinfo
  {author} {\bibfnamefont {M.}~\bibnamefont {Sato}},\ }\bibfield  {title}
  {\bibinfo {title} {Symmetry and topology in non-{H}ermitian physics},\ }\href
  {https://doi.org/10.1103/PhysRevX.9.041015} {\bibfield  {journal} {\bibinfo
  {journal} {Phys. Rev. X}\ }\textbf {\bibinfo {volume} {9}},\ \bibinfo {pages}
  {041015} (\bibinfo {year} {2019})}\BibitemShut {NoStop}%
\bibitem [{\citenamefont {Gong}\ \emph {et~al.}(2018)\citenamefont {Gong},
  \citenamefont {Ashida}, \citenamefont {Kawabata}, \citenamefont {Takasan},
  \citenamefont {Higashikawa},\ and\ \citenamefont {Ueda}}]{PhysRevX.8.031079}%
  \BibitemOpen
  \bibfield  {author} {\bibinfo {author} {\bibfnamefont {Z.}~\bibnamefont
  {Gong}}, \bibinfo {author} {\bibfnamefont {Y.}~\bibnamefont {Ashida}},
  \bibinfo {author} {\bibfnamefont {K.}~\bibnamefont {Kawabata}}, \bibinfo
  {author} {\bibfnamefont {K.}~\bibnamefont {Takasan}}, \bibinfo {author}
  {\bibfnamefont {S.}~\bibnamefont {Higashikawa}},\ and\ \bibinfo {author}
  {\bibfnamefont {M.}~\bibnamefont {Ueda}},\ }\bibfield  {title} {\bibinfo
  {title} {Topological phases of non-{H}ermitian systems},\ }\href
  {https://doi.org/10.1103/PhysRevX.8.031079} {\bibfield  {journal} {\bibinfo
  {journal} {Phys. Rev. X}\ }\textbf {\bibinfo {volume} {8}},\ \bibinfo {pages}
  {031079} (\bibinfo {year} {2018})}\BibitemShut {NoStop}%
\bibitem [{\citenamefont {Kobia\l{}ka}\ \emph {et~al.}(2024)\citenamefont
  {Kobia\l{}ka}, \citenamefont {Awoga}, \citenamefont {Leijnse}, \citenamefont
  {Doma\ifmmode~\acute{n}\else \'{n}\fi{}ski}, \citenamefont {Holmvall},\ and\
  \citenamefont {Black-Schaffer}}]{Kobialka2024Topological}%
  \BibitemOpen
  \bibfield  {author} {\bibinfo {author} {\bibfnamefont {A.}~\bibnamefont
  {Kobia\l{}ka}}, \bibinfo {author} {\bibfnamefont {O.~A.}\ \bibnamefont
  {Awoga}}, \bibinfo {author} {\bibfnamefont {M.}~\bibnamefont {Leijnse}},
  \bibinfo {author} {\bibfnamefont {T.}~\bibnamefont
  {Doma\ifmmode~\acute{n}\else \'{n}\fi{}ski}}, \bibinfo {author}
  {\bibfnamefont {P.}~\bibnamefont {Holmvall}},\ and\ \bibinfo {author}
  {\bibfnamefont {A.~M.}\ \bibnamefont {Black-Schaffer}},\ }\bibfield  {title}
  {\bibinfo {title} {Topological superconductivity in {F}ibonacci
  quasicrystal},\ }\href {https://doi.org/10.1103/PhysRevB.110.134508}
  {\bibfield  {journal} {\bibinfo  {journal} {Phys. Rev. B}\ }\textbf {\bibinfo
  {volume} {110}},\ \bibinfo {pages} {134508} (\bibinfo {year}
  {2024})}\BibitemShut {NoStop}%
\bibitem [{\citenamefont {Sticlet}\ \emph {et~al.}(2012)\citenamefont
  {Sticlet}, \citenamefont {Bena},\ and\ \citenamefont
  {Simon}}]{sticlet.bena.12}%
  \BibitemOpen
  \bibfield  {author} {\bibinfo {author} {\bibfnamefont {D.}~\bibnamefont
  {Sticlet}}, \bibinfo {author} {\bibfnamefont {C.}~\bibnamefont {Bena}},\ and\
  \bibinfo {author} {\bibfnamefont {P.}~\bibnamefont {Simon}},\ }\bibfield
  {title} {\bibinfo {title} {Spin and {M}ajorana polarization in topological
  superconducting wires},\ }\href
  {https://doi.org/10.1103/PhysRevLett.108.096802} {\bibfield  {journal}
  {\bibinfo  {journal} {Phys. Rev. Lett.}\ }\textbf {\bibinfo {volume} {108}},\
  \bibinfo {pages} {096802} (\bibinfo {year} {2012})}\BibitemShut {NoStop}%
\bibitem [{\citenamefont {Sedlmayr}\ and\ \citenamefont
  {Bena}(2015)}]{sedlmayr.bena2015}%
  \BibitemOpen
  \bibfield  {author} {\bibinfo {author} {\bibfnamefont {N.}~\bibnamefont
  {Sedlmayr}}\ and\ \bibinfo {author} {\bibfnamefont {C.}~\bibnamefont
  {Bena}},\ }\bibfield  {title} {\bibinfo {title} {Visualizing {M}ajorana bound
  states in one and two dimensions using the generalized {M}ajorana
  polarization},\ }\href {https://doi.org/10.1103/PhysRevB.92.115115}
  {\bibfield  {journal} {\bibinfo  {journal} {Phys. Rev. B}\ }\textbf {\bibinfo
  {volume} {92}},\ \bibinfo {pages} {115115} (\bibinfo {year}
  {2015})}\BibitemShut {NoStop}%
\bibitem [{\citenamefont {Sedlmayr}\ \emph {et~al.}(2016)\citenamefont
  {Sedlmayr}, \citenamefont {Aguiar-Hualde},\ and\ \citenamefont
  {Bena}}]{sedlmayr.aguiarhualde.16}%
  \BibitemOpen
  \bibfield  {author} {\bibinfo {author} {\bibfnamefont {N.}~\bibnamefont
  {Sedlmayr}}, \bibinfo {author} {\bibfnamefont {J.~M.}\ \bibnamefont
  {Aguiar-Hualde}},\ and\ \bibinfo {author} {\bibfnamefont {C.}~\bibnamefont
  {Bena}},\ }\bibfield  {title} {\bibinfo {title} {Majorana bound states in
  open quasi-one-dimensional and two-dimensional systems with transverse
  {R}ashba coupling},\ }\href {https://doi.org/10.1103/PhysRevB.93.155425}
  {\bibfield  {journal} {\bibinfo  {journal} {Phys. Rev. B}\ }\textbf {\bibinfo
  {volume} {93}},\ \bibinfo {pages} {155425} (\bibinfo {year}
  {2016})}\BibitemShut {NoStop}%
\bibitem [{\citenamefont {Bena}(2017)}]{Bena2017}%
  \BibitemOpen
  \bibfield  {author} {\bibinfo {author} {\bibfnamefont {C.}~\bibnamefont
  {Bena}},\ }\bibfield  {title} {\bibinfo {title} {Testing the formation of
  {Majorana} states using {Majorana} polarization},\ }\href
  {https://doi.org/10.1016/j.crhy.2017.09.005} {\bibfield  {journal} {\bibinfo
  {journal} {Comptes Rendus. Physique}\ }\textbf {\bibinfo {volume} {18}},\
  \bibinfo {pages} {349} (\bibinfo {year} {2017})}\BibitemShut {NoStop}%
\bibitem [{\citenamefont {Sedlmayr}\ \emph {et~al.}(2017)\citenamefont
  {Sedlmayr}, \citenamefont {Kaladzhyan}, \citenamefont {Dutreix},\ and\
  \citenamefont {Bena}}]{Sedlmayr.Kaladzhyan.2017.Bulk}%
  \BibitemOpen
  \bibfield  {author} {\bibinfo {author} {\bibfnamefont {N.}~\bibnamefont
  {Sedlmayr}}, \bibinfo {author} {\bibfnamefont {V.}~\bibnamefont
  {Kaladzhyan}}, \bibinfo {author} {\bibfnamefont {C.}~\bibnamefont
  {Dutreix}},\ and\ \bibinfo {author} {\bibfnamefont {C.}~\bibnamefont
  {Bena}},\ }\bibfield  {title} {\bibinfo {title} {Bulk boundary correspondence
  and the existence of {M}ajorana bound states on the edges of 2d topological
  superconductors},\ }\href {https://doi.org/10.1103/PhysRevB.96.184516}
  {\bibfield  {journal} {\bibinfo  {journal} {Phys. Rev. B}\ }\textbf {\bibinfo
  {volume} {96}},\ \bibinfo {pages} {184516} (\bibinfo {year}
  {2017})}\BibitemShut {NoStop}%
\bibitem [{\citenamefont {Awoga}\ \emph {et~al.}(2024)\citenamefont {Awoga},
  \citenamefont {Ioannidis}, \citenamefont {Mishra}, \citenamefont {Leijnse},
  \citenamefont {Trif},\ and\ \citenamefont
  {Posske}}]{PhysRevResearch.6.033154}%
  \BibitemOpen
  \bibfield  {author} {\bibinfo {author} {\bibfnamefont {O.~A.}\ \bibnamefont
  {Awoga}}, \bibinfo {author} {\bibfnamefont {I.}~\bibnamefont {Ioannidis}},
  \bibinfo {author} {\bibfnamefont {A.}~\bibnamefont {Mishra}}, \bibinfo
  {author} {\bibfnamefont {M.}~\bibnamefont {Leijnse}}, \bibinfo {author}
  {\bibfnamefont {M.}~\bibnamefont {Trif}},\ and\ \bibinfo {author}
  {\bibfnamefont {T.}~\bibnamefont {Posske}},\ }\bibfield  {title} {\bibinfo
  {title} {Controlling {M}ajorana hybridization in magnetic
  chain-superconductor systems},\ }\href
  {https://doi.org/10.1103/PhysRevResearch.6.033154} {\bibfield  {journal}
  {\bibinfo  {journal} {Phys. Rev. Res.}\ }\textbf {\bibinfo {volume} {6}},\
  \bibinfo {pages} {033154} (\bibinfo {year} {2024})}\BibitemShut {NoStop}%
\bibitem [{\citenamefont {Souto}\ \emph {et~al.}(2023)\citenamefont {Souto},
  \citenamefont {Tsintzis}, \citenamefont {Leijnse},\ and\ \citenamefont
  {Danon}}]{Souto2024Probing}%
  \BibitemOpen
  \bibfield  {author} {\bibinfo {author} {\bibfnamefont {R.~S.}\ \bibnamefont
  {Souto}}, \bibinfo {author} {\bibfnamefont {A.}~\bibnamefont {Tsintzis}},
  \bibinfo {author} {\bibfnamefont {M.}~\bibnamefont {Leijnse}},\ and\ \bibinfo
  {author} {\bibfnamefont {J.}~\bibnamefont {Danon}},\ }\bibfield  {title}
  {\bibinfo {title} {Probing {M}ajorana localization in minimal {K}itaev chains
  through a quantum dot},\ }\href
  {https://doi.org/10.1103/PhysRevResearch.5.043182} {\bibfield  {journal}
  {\bibinfo  {journal} {Phys. Rev. Res.}\ }\textbf {\bibinfo {volume} {5}},\
  \bibinfo {pages} {043182} (\bibinfo {year} {2023})}\BibitemShut {NoStop}%
\bibitem [{\citenamefont {Samuelson}\ \emph {et~al.}(2024)\citenamefont
  {Samuelson}, \citenamefont {Svensson},\ and\ \citenamefont
  {Leijnse}}]{Samuelson2024Minimal}%
  \BibitemOpen
  \bibfield  {author} {\bibinfo {author} {\bibfnamefont {W.}~\bibnamefont
  {Samuelson}}, \bibinfo {author} {\bibfnamefont {V.}~\bibnamefont
  {Svensson}},\ and\ \bibinfo {author} {\bibfnamefont {M.}~\bibnamefont
  {Leijnse}},\ }\bibfield  {title} {\bibinfo {title} {Minimal quantum dot based
  {K}itaev chain with only local superconducting proximity effect},\ }\href
  {https://doi.org/10.1103/PhysRevB.109.035415} {\bibfield  {journal} {\bibinfo
   {journal} {Phys. Rev. B}\ }\textbf {\bibinfo {volume} {109}},\ \bibinfo
  {pages} {035415} (\bibinfo {year} {2024})}\BibitemShut {NoStop}%
\bibitem [{\citenamefont {He}\ \emph {et~al.}(2025)\citenamefont {He},
  \citenamefont {Bi}, \citenamefont {Fan}, \citenamefont {Wang}, \citenamefont
  {Li},\ and\ \citenamefont {Xu}}]{He.et.al.2025.Manipulating}%
  \BibitemOpen
  \bibfield  {author} {\bibinfo {author} {\bibfnamefont {S.-R.}\ \bibnamefont
  {He}}, \bibinfo {author} {\bibfnamefont {X.-L.}\ \bibnamefont {Bi}}, \bibinfo
  {author} {\bibfnamefont {J.}~\bibnamefont {Fan}}, \bibinfo {author}
  {\bibfnamefont {Z.-H.}\ \bibnamefont {Wang}}, \bibinfo {author}
  {\bibfnamefont {L.}~\bibnamefont {Li}},\ and\ \bibinfo {author}
  {\bibfnamefont {D.-H.}\ \bibnamefont {Xu}},\ }\bibfield  {title} {\bibinfo
  {title} {Manipulating and braiding {M}ajorana corner modes on a agome
  lattice},\ }\href {https://doi.org/10.1103/PhysRevB.111.144506} {\bibfield
  {journal} {\bibinfo  {journal} {Phys. Rev. B}\ }\textbf {\bibinfo {volume}
  {111}},\ \bibinfo {pages} {144506} (\bibinfo {year} {2025})}\BibitemShut
  {NoStop}%
\bibitem [{\citenamefont {Karoliya}\ \emph {et~al.}()\citenamefont {Karoliya},
  \citenamefont {Tewari},\ and\ \citenamefont {Sharma}}]{karoliya2025majorana}%
  \BibitemOpen
  \bibfield  {author} {\bibinfo {author} {\bibfnamefont {S.}~\bibnamefont
  {Karoliya}}, \bibinfo {author} {\bibfnamefont {S.}~\bibnamefont {Tewari}},\
  and\ \bibinfo {author} {\bibfnamefont {G.}~\bibnamefont {Sharma}},\
  }\bibfield  {title} {\bibinfo {title} {Majorana modes in graphene strips:
  polarization, wavefunctions, disorder, and {A}ndreev states},\ }\href@noop {}
  {\bibinfo  {journal} {ArXiv}\ }\BibitemShut {NoStop}%
\bibitem [{Note1()}]{Note1}%
  \BibitemOpen
\bibfield  {journal} {  }\bibinfo {note} {We note that the MP has already been
  studied in NH systems \cite {Wang.2021.Majorana} but still as a local spatial
  topological indicator. This leaves the nonlocal MP in NH systems
  unexplored.}\BibitemShut {Stop}%
\bibitem [{\citenamefont {Moiseyev}(2011)}]{Moiseyev}%
  \BibitemOpen
  \bibfield  {author} {\bibinfo {author} {\bibfnamefont {N.}~\bibnamefont
  {Moiseyev}},\ }\href@noop {} {\emph {\bibinfo {title} {Non-{H}ermitian
  Quantum Mechanics}}}\ (\bibinfo  {publisher} {Cambridge University Press},\
  \bibinfo {year} {2011})\BibitemShut {NoStop}%
\bibitem [{\citenamefont {Kunst}\ \emph {et~al.}(2018)\citenamefont {Kunst},
  \citenamefont {Edvardsson}, \citenamefont {Budich},\ and\ \citenamefont
  {Bergholtz}}]{Kunst.2018.Biorthogonal}%
  \BibitemOpen
  \bibfield  {author} {\bibinfo {author} {\bibfnamefont {F.~K.}\ \bibnamefont
  {Kunst}}, \bibinfo {author} {\bibfnamefont {E.}~\bibnamefont {Edvardsson}},
  \bibinfo {author} {\bibfnamefont {J.~C.}\ \bibnamefont {Budich}},\ and\
  \bibinfo {author} {\bibfnamefont {E.~J.}\ \bibnamefont {Bergholtz}},\
  }\bibfield  {title} {\bibinfo {title} {Biorthogonal bulk-boundary
  correspondence in non-{H}ermitian systems},\ }\href
  {https://doi.org/10.1103/PhysRevLett.121.026808} {\bibfield  {journal}
  {\bibinfo  {journal} {Phys. Rev. Lett.}\ }\textbf {\bibinfo {volume} {121}},\
  \bibinfo {pages} {026808} (\bibinfo {year} {2018})}\BibitemShut {NoStop}%
\bibitem [{Note2()}]{Note2}%
  \BibitemOpen
  \bibinfo {note} {Another scenario with $\protect \mathcal R =0$ occurs when
  ${\protect \mathcal {P}^{\protect \rm RR}}= {\protect \mathcal {P}^{\protect
  \rm LL}}= {\protect \mathcal {P}^{\protect \rm LR}}={\protect \mathcal
  {P}^{\protect \rm RL}}= 0$ for any $\Gamma $. This is a trivial solution
  since the superconductor does not contain MZMs or EPs in this
  case.}\BibitemShut {Stop}%
\bibitem [{\citenamefont {Trefethen}\ and\ \citenamefont
  {Embree}(2005)}]{trefethen2005spectra}%
  \BibitemOpen
  \bibfield  {author} {\bibinfo {author} {\bibfnamefont {L.~N.}\ \bibnamefont
  {Trefethen}}\ and\ \bibinfo {author} {\bibfnamefont {M.}~\bibnamefont
  {Embree}},\ }\href@noop {} {\emph {\bibinfo {title} {Spectra and
  Pseudospectra: {T}he Behavior of Nonnormal Matrices and Operators}}}\
  (\bibinfo  {publisher} {Princeton University Press},\ \bibinfo {year}
  {2005})\BibitemShut {NoStop}%
\bibitem [{\citenamefont {Hatano}\ and\ \citenamefont
  {Nelson}(1996)}]{Hatano.1996.Localization}%
  \BibitemOpen
  \bibfield  {author} {\bibinfo {author} {\bibfnamefont {N.}~\bibnamefont
  {Hatano}}\ and\ \bibinfo {author} {\bibfnamefont {D.~R.}\ \bibnamefont
  {Nelson}},\ }\bibfield  {title} {\bibinfo {title} {Localization transitions
  in non-{H}ermitian quantum mechanics},\ }\href
  {https://doi.org/10.1103/PhysRevLett.77.570} {\bibfield  {journal} {\bibinfo
  {journal} {Phys. Rev. Lett.}\ }\textbf {\bibinfo {volume} {77}},\ \bibinfo
  {pages} {570} (\bibinfo {year} {1996})}\BibitemShut {NoStop}%
\bibitem [{\citenamefont {Ekman}\ \emph {et~al.}(2026)\citenamefont {Ekman},
  \citenamefont {Bergholtz},\ and\ \citenamefont
  {Molignini}}]{ekman2026symmetry}%
  \BibitemOpen
  \bibfield  {author} {\bibinfo {author} {\bibfnamefont {C.}~\bibnamefont
  {Ekman}}, \bibinfo {author} {\bibfnamefont {E.~J.}\ \bibnamefont
  {Bergholtz}},\ and\ \bibinfo {author} {\bibfnamefont {P.}~\bibnamefont
  {Molignini}},\ }\bibfield  {title} {\bibinfo {title} {Symmetry-fractionalized
  skin effects in non-{H}ermitian {L}uttinger liquids},\ }\href
  {https://arxiv.org/abs/2603.28849} {\bibfield  {journal} {\bibinfo  {journal}
  {arXiv}\ }\textbf {\bibinfo {volume} {2603.28849}} (\bibinfo {year}
  {2026})}\BibitemShut {NoStop}%
\bibitem [{Note3()}]{Note3}%
  \BibitemOpen
  \bibinfo {note} {This is applicable to non-Hermitian Kitaev models with other
  types of non-Hermiticity~\cite
  {PhysRevX.9.041015,Okuma2023,cayao2023nonhermitian}}\BibitemShut {NoStop}%
\bibitem [{\citenamefont {Leijnse}\ and\ \citenamefont
  {Flensberg}(2012)}]{Leijnse2012Parity}%
  \BibitemOpen
  \bibfield  {author} {\bibinfo {author} {\bibfnamefont {M.}~\bibnamefont
  {Leijnse}}\ and\ \bibinfo {author} {\bibfnamefont {K.}~\bibnamefont
  {Flensberg}},\ }\bibfield  {title} {\bibinfo {title} {Parity qubits and poor
  man's {M}ajorana bound states in double quantum dots},\ }\href
  {https://doi.org/10.1103/PhysRevB.86.134528} {\bibfield  {journal} {\bibinfo
  {journal} {Phys. Rev. B}\ }\textbf {\bibinfo {volume} {86}},\ \bibinfo
  {pages} {134528} (\bibinfo {year} {2012})}\BibitemShut {NoStop}%
\bibitem [{\citenamefont {Sau}\ and\ \citenamefont {Sarma}(2012)}]{Sau2012}%
  \BibitemOpen
  \bibfield  {author} {\bibinfo {author} {\bibfnamefont {J.~D.}\ \bibnamefont
  {Sau}}\ and\ \bibinfo {author} {\bibfnamefont {S.~D.}\ \bibnamefont
  {Sarma}},\ }\bibfield  {title} {\bibinfo {title} {Realizing a robust
  practical {M}ajorana chain in a quantum-dot-superconductor linear array},\
  }\href {https://doi.org/10.1038/ncomms1966} {\bibfield  {journal} {\bibinfo
  {journal} {Nat. Commun.}\ }\textbf {\bibinfo {volume} {3}},\ \bibinfo {pages}
  {964} (\bibinfo {year} {2012})}\BibitemShut {NoStop}%
\bibitem [{\citenamefont {Cayao}(2024{\natexlab{b}})}]{PhysRevB.110.125408}%
  \BibitemOpen
  \bibfield  {author} {\bibinfo {author} {\bibfnamefont {J.}~\bibnamefont
  {Cayao}},\ }\bibfield  {title} {\bibinfo {title} {Emergent pair symmetries in
  systems with poor man's {M}ajorana modes},\ }\href
  {https://doi.org/10.1103/PhysRevB.110.125408} {\bibfield  {journal} {\bibinfo
   {journal} {Phys. Rev. B}\ }\textbf {\bibinfo {volume} {110}},\ \bibinfo
  {pages} {125408} (\bibinfo {year} {2024}{\natexlab{b}})}\BibitemShut
  {NoStop}%
\bibitem [{\citenamefont {Tsintzis}\ \emph {et~al.}(2022)\citenamefont
  {Tsintzis}, \citenamefont {Souto},\ and\ \citenamefont
  {Leijnse}}]{Tsintzis2022Creating}%
  \BibitemOpen
  \bibfield  {author} {\bibinfo {author} {\bibfnamefont {A.}~\bibnamefont
  {Tsintzis}}, \bibinfo {author} {\bibfnamefont {R.~S.}\ \bibnamefont
  {Souto}},\ and\ \bibinfo {author} {\bibfnamefont {M.}~\bibnamefont
  {Leijnse}},\ }\bibfield  {title} {\bibinfo {title} {Creating and detecting
  poor man's {M}ajorana bound states in interacting quantum dots},\ }\href
  {https://doi.org/10.1103/PhysRevB.106.L201404} {\bibfield  {journal}
  {\bibinfo  {journal} {Phys. Rev. B}\ }\textbf {\bibinfo {volume} {106}},\
  \bibinfo {pages} {L201404} (\bibinfo {year} {2022})}\BibitemShut {NoStop}%
\bibitem [{\citenamefont {Alvarado}\ \emph {et~al.}(2024)\citenamefont
  {Alvarado}, \citenamefont {Yeyati}, \citenamefont {Aguado},\ and\
  \citenamefont {Souto}}]{PhysRevB.110.245144}%
  \BibitemOpen
  \bibfield  {author} {\bibinfo {author} {\bibfnamefont {M.}~\bibnamefont
  {Alvarado}}, \bibinfo {author} {\bibfnamefont {A.~L.}\ \bibnamefont
  {Yeyati}}, \bibinfo {author} {\bibfnamefont {R.}~\bibnamefont {Aguado}},\
  and\ \bibinfo {author} {\bibfnamefont {R.~S.}\ \bibnamefont {Souto}},\
  }\bibfield  {title} {\bibinfo {title} {Interplay between {M}ajorana and
  {S}hiba states in a minimal {K}itaev chain coupled to a superconductor},\
  }\href {https://doi.org/10.1103/PhysRevB.110.245144} {\bibfield  {journal}
  {\bibinfo  {journal} {Phys. Rev. B}\ }\textbf {\bibinfo {volume} {110}},\
  \bibinfo {pages} {245144} (\bibinfo {year} {2024})}\BibitemShut {NoStop}%
\bibitem [{\citenamefont {Luethi}\ \emph {et~al.}(2024)\citenamefont {Luethi},
  \citenamefont {Legg}, \citenamefont {Loss},\ and\ \citenamefont
  {Klinovaja}}]{PhysRevB.110.245412}%
  \BibitemOpen
  \bibfield  {author} {\bibinfo {author} {\bibfnamefont {M.}~\bibnamefont
  {Luethi}}, \bibinfo {author} {\bibfnamefont {H.~F.}\ \bibnamefont {Legg}},
  \bibinfo {author} {\bibfnamefont {D.}~\bibnamefont {Loss}},\ and\ \bibinfo
  {author} {\bibfnamefont {J.}~\bibnamefont {Klinovaja}},\ }\bibfield  {title}
  {\bibinfo {title} {From perfect to imperfect poor man's {M}ajoranas in
  minimal {K}itaev chains},\ }\href
  {https://doi.org/10.1103/PhysRevB.110.245412} {\bibfield  {journal} {\bibinfo
   {journal} {Phys. Rev. B}\ }\textbf {\bibinfo {volume} {110}},\ \bibinfo
  {pages} {245412} (\bibinfo {year} {2024})}\BibitemShut {NoStop}%
\bibitem [{\citenamefont {Kotetes}\ \emph {et~al.}(2024)\citenamefont
  {Kotetes}, \citenamefont {Roig},\ and\ \citenamefont
  {Andersen}}]{kotetes2024nonRecifourpi}%
  \BibitemOpen
  \bibfield  {author} {\bibinfo {author} {\bibfnamefont {P.}~\bibnamefont
  {Kotetes}}, \bibinfo {author} {\bibfnamefont {M.}~\bibnamefont {Roig}},\ and\
  \bibinfo {author} {\bibfnamefont {B.~M.}\ \bibnamefont {Andersen}},\
  }\bibfield  {title} {\bibinfo {title} {Nonreciprocal equilibrium
  4$\pi$-periodic {J}osephson effect from poor man's {M}ajorana zero modes},\
  }\href {https://arxiv.org/abs/2409.13027} {\bibfield  {journal} {\bibinfo
  {journal} {arXiv:2409.13027}\ } (\bibinfo {year} {2024})}\BibitemShut
  {NoStop}%
\bibitem [{\citenamefont {Vimal}\ and\ \citenamefont
  {Cayao}(2025)}]{vimal2025EntMKC}%
  \BibitemOpen
  \bibfield  {author} {\bibinfo {author} {\bibfnamefont {V.~K.}\ \bibnamefont
  {Vimal}}\ and\ \bibinfo {author} {\bibfnamefont {J.}~\bibnamefont {Cayao}},\
  }\bibfield  {title} {\bibinfo {title} {Entanglement dynamics in minimal
  {K}itaev chains},\ }\href {https://arxiv.org/abs/2507.17586} {\bibfield
  {journal} {\bibinfo  {journal} {arXiv: 2507.17586}\ } (\bibinfo {year}
  {2025})}\BibitemShut {NoStop}%
\bibitem [{\citenamefont {Luethi}\ \emph {et~al.}(2025)\citenamefont {Luethi},
  \citenamefont {Legg}, \citenamefont {Loss},\ and\ \citenamefont
  {Klinovaja}}]{PhysRevB.111.115419}%
  \BibitemOpen
  \bibfield  {author} {\bibinfo {author} {\bibfnamefont {M.}~\bibnamefont
  {Luethi}}, \bibinfo {author} {\bibfnamefont {H.~F.}\ \bibnamefont {Legg}},
  \bibinfo {author} {\bibfnamefont {D.}~\bibnamefont {Loss}},\ and\ \bibinfo
  {author} {\bibfnamefont {J.}~\bibnamefont {Klinovaja}},\ }\bibfield  {title}
  {\bibinfo {title} {Fate of poor man's {M}ajoranas in the long {K}itaev chain
  limit},\ }\href {https://doi.org/10.1103/PhysRevB.111.115419} {\bibfield
  {journal} {\bibinfo  {journal} {Phys. Rev. B}\ }\textbf {\bibinfo {volume}
  {111}},\ \bibinfo {pages} {115419} (\bibinfo {year} {2025})}\BibitemShut
  {NoStop}%
\bibitem [{\citenamefont {Cayao}\ and\ \citenamefont
  {Sato}(2026{\natexlab{b}})}]{cayao2026nonlocal}%
  \BibitemOpen
  \bibfield  {author} {\bibinfo {author} {\bibfnamefont {J.}~\bibnamefont
  {Cayao}}\ and\ \bibinfo {author} {\bibfnamefont {M.}~\bibnamefont {Sato}},\
  }\bibfield  {title} {\bibinfo {title} {Nonlocal {J}osephson diode effect in
  minimal {K}itaev chains},\ }\href {https://doi.org/10.1103/hf7s-f7tj}
  {\bibfield  {journal} {\bibinfo  {journal} {Phys. Rev. Res.}\ }\textbf
  {\bibinfo {volume} {8}},\ \bibinfo {pages} {013326} (\bibinfo {year}
  {2026}{\natexlab{b}})}\BibitemShut {NoStop}%
\bibitem [{\citenamefont {Cayao}\ and\ \citenamefont
  {Sato}(2026{\natexlab{c}})}]{cayao2026AEPMKC}%
  \BibitemOpen
  \bibfield  {author} {\bibinfo {author} {\bibfnamefont {J.}~\bibnamefont
  {Cayao}}\ and\ \bibinfo {author} {\bibfnamefont {M.}~\bibnamefont {Sato}},\
  }\bibfield  {title} {\bibinfo {title} {Andreev exceptional points in
  {J}osephson junctions formed by minimal {K}itaev chains},\ }\href
  {https://arxiv.org/abs/2606.23956} {\bibfield  {journal} {\bibinfo  {journal}
  {arXiv:2606.23956}\ } (\bibinfo {year} {2026}{\natexlab{c}})}\BibitemShut
  {NoStop}%
\bibitem [{\citenamefont {Dvir}\ \emph {et~al.}(2023)\citenamefont {Dvir},
  \citenamefont {Wang}, \citenamefont {van Loo}, \citenamefont {Liu},
  \citenamefont {Mazur}, \citenamefont {Bordin}, \citenamefont {Ten~Haaf},
  \citenamefont {Wang}, \citenamefont {van Driel}, \citenamefont {Zatelli},
  \citenamefont {Li}, \citenamefont {Malinowski}, \citenamefont {Gazibegovic},
  \citenamefont {Badawy}, \citenamefont {Bakkers}, \citenamefont {Wimmer},\
  and\ \citenamefont {Kouwenhoven}}]{dvir2023realization}%
  \BibitemOpen
  \bibfield  {author} {\bibinfo {author} {\bibfnamefont {T.}~\bibnamefont
  {Dvir}}, \bibinfo {author} {\bibfnamefont {G.}~\bibnamefont {Wang}}, \bibinfo
  {author} {\bibfnamefont {N.}~\bibnamefont {van Loo}}, \bibinfo {author}
  {\bibfnamefont {C.-X.}\ \bibnamefont {Liu}}, \bibinfo {author} {\bibfnamefont
  {G.~P.}\ \bibnamefont {Mazur}}, \bibinfo {author} {\bibfnamefont
  {A.}~\bibnamefont {Bordin}}, \bibinfo {author} {\bibfnamefont {S.~L.}\
  \bibnamefont {Ten~Haaf}}, \bibinfo {author} {\bibfnamefont {J.-Y.}\
  \bibnamefont {Wang}}, \bibinfo {author} {\bibfnamefont {D.}~\bibnamefont {van
  Driel}}, \bibinfo {author} {\bibfnamefont {F.}~\bibnamefont {Zatelli}},
  \bibinfo {author} {\bibfnamefont {X.}~\bibnamefont {Li}}, \bibinfo {author}
  {\bibfnamefont {F.~K.}\ \bibnamefont {Malinowski}}, \bibinfo {author}
  {\bibfnamefont {S.}~\bibnamefont {Gazibegovic}}, \bibinfo {author}
  {\bibfnamefont {G.}~\bibnamefont {Badawy}}, \bibinfo {author} {\bibfnamefont
  {E.~P. A.~M.}\ \bibnamefont {Bakkers}}, \bibinfo {author} {\bibfnamefont
  {M.}~\bibnamefont {Wimmer}},\ and\ \bibinfo {author} {\bibfnamefont {L.~P.}\
  \bibnamefont {Kouwenhoven}},\ }\bibfield  {title} {\bibinfo {title}
  {Realization of a minimal {K}itaev chain in coupled quantum dots},\ }\href
  {https://doi.org/10.1038/s41586-022-05585-1} {\bibfield  {journal} {\bibinfo
  {journal} {Nature}\ }\textbf {\bibinfo {volume} {614}},\ \bibinfo {pages}
  {445} (\bibinfo {year} {2023})}\BibitemShut {NoStop}%
\bibitem [{\citenamefont {ten Haaf}\ \emph {et~al.}(2024)\citenamefont {ten
  Haaf}, \citenamefont {Wang}, \citenamefont {Bozkurt}, \citenamefont {Liu},
  \citenamefont {Kulesh}, \citenamefont {Kim}, \citenamefont {Xiao},
  \citenamefont {Thomas}, \citenamefont {Manfra}, \citenamefont {Dvir},
  \citenamefont {Wimmer},\ and\ \citenamefont {Goswami}}]{Haaf2024}%
  \BibitemOpen
  \bibfield  {author} {\bibinfo {author} {\bibfnamefont {S.~L.~D.}\
  \bibnamefont {ten Haaf}}, \bibinfo {author} {\bibfnamefont {Q.}~\bibnamefont
  {Wang}}, \bibinfo {author} {\bibfnamefont {A.~M.}\ \bibnamefont {Bozkurt}},
  \bibinfo {author} {\bibfnamefont {C.-X.}\ \bibnamefont {Liu}}, \bibinfo
  {author} {\bibfnamefont {I.}~\bibnamefont {Kulesh}}, \bibinfo {author}
  {\bibfnamefont {P.}~\bibnamefont {Kim}}, \bibinfo {author} {\bibfnamefont
  {D.}~\bibnamefont {Xiao}}, \bibinfo {author} {\bibfnamefont {C.}~\bibnamefont
  {Thomas}}, \bibinfo {author} {\bibfnamefont {M.~J.}\ \bibnamefont {Manfra}},
  \bibinfo {author} {\bibfnamefont {T.}~\bibnamefont {Dvir}}, \bibinfo {author}
  {\bibfnamefont {M.}~\bibnamefont {Wimmer}},\ and\ \bibinfo {author}
  {\bibfnamefont {S.}~\bibnamefont {Goswami}},\ }\bibfield  {title} {\bibinfo
  {title} {A two-site {K}itaev chain in a two-dimensional electron gas},\
  }\href {https://doi.org/10.1038/s41586-024-07434-9} {\bibfield  {journal}
  {\bibinfo  {journal} {Nature}\ }\textbf {\bibinfo {volume} {630}},\ \bibinfo
  {pages} {329} (\bibinfo {year} {2024})}\BibitemShut {NoStop}%
\bibitem [{\citenamefont {Zatelli}\ \emph {et~al.}(2024)\citenamefont
  {Zatelli}, \citenamefont {van Driel}, \citenamefont {Xu}, \citenamefont
  {Wang}, \citenamefont {Liu}, \citenamefont {Bordin}, \citenamefont {Roovers},
  \citenamefont {Mazur}, \citenamefont {van Loo}, \citenamefont {Wolff},
  \citenamefont {Bozkurt}, \citenamefont {Badawy}, \citenamefont {Gazibegovic},
  \citenamefont {Bakkers}, \citenamefont {Wimmer}, \citenamefont
  {Kouwenhoven},\ and\ \citenamefont {Dvir}}]{Zatelli_2024}%
  \BibitemOpen
  \bibfield  {author} {\bibinfo {author} {\bibfnamefont {F.}~\bibnamefont
  {Zatelli}}, \bibinfo {author} {\bibfnamefont {D.}~\bibnamefont {van Driel}},
  \bibinfo {author} {\bibfnamefont {D.}~\bibnamefont {Xu}}, \bibinfo {author}
  {\bibfnamefont {G.}~\bibnamefont {Wang}}, \bibinfo {author} {\bibfnamefont
  {C.-X.}\ \bibnamefont {Liu}}, \bibinfo {author} {\bibfnamefont
  {A.}~\bibnamefont {Bordin}}, \bibinfo {author} {\bibfnamefont
  {B.}~\bibnamefont {Roovers}}, \bibinfo {author} {\bibfnamefont {G.~P.}\
  \bibnamefont {Mazur}}, \bibinfo {author} {\bibfnamefont {N.}~\bibnamefont
  {van Loo}}, \bibinfo {author} {\bibfnamefont {J.~C.}\ \bibnamefont {Wolff}},
  \bibinfo {author} {\bibfnamefont {A.~M.}\ \bibnamefont {Bozkurt}}, \bibinfo
  {author} {\bibfnamefont {G.}~\bibnamefont {Badawy}}, \bibinfo {author}
  {\bibfnamefont {S.}~\bibnamefont {Gazibegovic}}, \bibinfo {author}
  {\bibfnamefont {E.~P. A.~M.}\ \bibnamefont {Bakkers}}, \bibinfo {author}
  {\bibfnamefont {M.}~\bibnamefont {Wimmer}}, \bibinfo {author} {\bibfnamefont
  {L.~P.}\ \bibnamefont {Kouwenhoven}},\ and\ \bibinfo {author} {\bibfnamefont
  {T.}~\bibnamefont {Dvir}},\ }\bibfield  {title} {\bibinfo {title} {Robust
  poor man’s {M}ajorana zero modes using {Yu-Shiba-Rusinov} states},\ }\href
  {http://dx.doi.org/10.1038/s41467-024-52066-2} {\bibfield  {journal}
  {\bibinfo  {journal} {Nat. Commun.}\ }\textbf {\bibinfo {volume} {15}},\
  \bibinfo {pages} {7933} (\bibinfo {year} {2024})}\BibitemShut {NoStop}%
\bibitem [{\citenamefont {Bordin}\ \emph {et~al.}(2025)\citenamefont {Bordin},
  \citenamefont {Liu}, \citenamefont {Dvir}, \citenamefont {Zatelli},
  \citenamefont {Ten~Haaf}, \citenamefont {van Driel}, \citenamefont {Wang},
  \citenamefont {Van~Loo}, \citenamefont {Zhang}, \citenamefont {Wolff},
  \citenamefont {Caekenberghe}, \citenamefont {Badawy}, \citenamefont
  {Gazibegovic}, \citenamefont {Bakkers}, \citenamefont {Wimmer}, \citenamefont
  {Kouwenhoven},\ and\ \citenamefont {Mazur}}]{bordin2025enhanced}%
  \BibitemOpen
  \bibfield  {author} {\bibinfo {author} {\bibfnamefont {A.}~\bibnamefont
  {Bordin}}, \bibinfo {author} {\bibfnamefont {C.-X.}\ \bibnamefont {Liu}},
  \bibinfo {author} {\bibfnamefont {T.}~\bibnamefont {Dvir}}, \bibinfo {author}
  {\bibfnamefont {F.}~\bibnamefont {Zatelli}}, \bibinfo {author} {\bibfnamefont
  {S.~L.}\ \bibnamefont {Ten~Haaf}}, \bibinfo {author} {\bibfnamefont
  {D.}~\bibnamefont {van Driel}}, \bibinfo {author} {\bibfnamefont
  {G.}~\bibnamefont {Wang}}, \bibinfo {author} {\bibfnamefont {N.}~\bibnamefont
  {Van~Loo}}, \bibinfo {author} {\bibfnamefont {Y.}~\bibnamefont {Zhang}},
  \bibinfo {author} {\bibfnamefont {J.~C.}\ \bibnamefont {Wolff}}, \bibinfo
  {author} {\bibfnamefont {T.~V.}\ \bibnamefont {Caekenberghe}}, \bibinfo
  {author} {\bibfnamefont {G.}~\bibnamefont {Badawy}}, \bibinfo {author}
  {\bibfnamefont {S.}~\bibnamefont {Gazibegovic}}, \bibinfo {author}
  {\bibfnamefont {E.~P. A.~M.}\ \bibnamefont {Bakkers}}, \bibinfo {author}
  {\bibfnamefont {M.}~\bibnamefont {Wimmer}}, \bibinfo {author} {\bibfnamefont
  {L.~P.}\ \bibnamefont {Kouwenhoven}},\ and\ \bibinfo {author} {\bibfnamefont
  {G.~P.}\ \bibnamefont {Mazur}},\ }\bibfield  {title} {\bibinfo {title}
  {Enhanced {M}ajorana stability in a three-site {K}itaev chain},\ }\href
  {https://doi.org/10.1038/s41565-025-01894-4} {\bibfield  {journal} {\bibinfo
  {journal} {Nat. Nanotech.}\ }\textbf {\bibinfo {volume} {20}},\ \bibinfo
  {pages} {726} (\bibinfo {year} {2025})}\BibitemShut {NoStop}%
\bibitem [{\citenamefont {Kulesh}\ \emph {et~al.}(2025)\citenamefont {Kulesh},
  \citenamefont {ten Haaf}, \citenamefont {Wang}, \citenamefont {Sietses},
  \citenamefont {Zhang}, \citenamefont {Roelofs}, \citenamefont {Prosko},
  \citenamefont {Xiao}, \citenamefont {Thomas}, \citenamefont {Manfra},\ and\
  \citenamefont {Goswami}}]{Kulesh.2025.Flux}%
  \BibitemOpen
  \bibfield  {author} {\bibinfo {author} {\bibfnamefont {I.}~\bibnamefont
  {Kulesh}}, \bibinfo {author} {\bibfnamefont {S.~L.~D.}\ \bibnamefont {ten
  Haaf}}, \bibinfo {author} {\bibfnamefont {Q.}~\bibnamefont {Wang}}, \bibinfo
  {author} {\bibfnamefont {V.~P.~M.}\ \bibnamefont {Sietses}}, \bibinfo
  {author} {\bibfnamefont {Y.}~\bibnamefont {Zhang}}, \bibinfo {author}
  {\bibfnamefont {S.~R.}\ \bibnamefont {Roelofs}}, \bibinfo {author}
  {\bibfnamefont {C.~G.}\ \bibnamefont {Prosko}}, \bibinfo {author}
  {\bibfnamefont {D.}~\bibnamefont {Xiao}}, \bibinfo {author} {\bibfnamefont
  {C.}~\bibnamefont {Thomas}}, \bibinfo {author} {\bibfnamefont {M.~J.}\
  \bibnamefont {Manfra}},\ and\ \bibinfo {author} {\bibfnamefont
  {S.}~\bibnamefont {Goswami}},\ }\bibfield  {title} {\bibinfo {title}
  {Flux-controlled two-site {K}itaev chain},\ }\href
  {https://doi.org/10.1103/r9pv-2prs} {\bibfield  {journal} {\bibinfo
  {journal} {Phys. Rev. Lett.}\ }\textbf {\bibinfo {volume} {135}},\ \bibinfo
  {pages} {056301} (\bibinfo {year} {2025})}\BibitemShut {NoStop}%
\bibitem [{\citenamefont {Dourado}\ \emph {et~al.}(2025)\citenamefont
  {Dourado}, \citenamefont {Leijnse},\ and\ \citenamefont
  {Souto}}]{Dourado.2025.Majorana}%
  \BibitemOpen
  \bibfield  {author} {\bibinfo {author} {\bibfnamefont {R.~A.}\ \bibnamefont
  {Dourado}}, \bibinfo {author} {\bibfnamefont {M.}~\bibnamefont {Leijnse}},\
  and\ \bibinfo {author} {\bibfnamefont {R.~S.}\ \bibnamefont {Souto}},\
  }\bibfield  {title} {\bibinfo {title} {Majorana sweet spots in three-site
  {K}itaev chains},\ }\href {https://doi.org/10.1103/PhysRevB.111.235409}
  {\bibfield  {journal} {\bibinfo  {journal} {Phys. Rev. B}\ }\textbf {\bibinfo
  {volume} {111}},\ \bibinfo {pages} {235409} (\bibinfo {year}
  {2025})}\BibitemShut {NoStop}%
\bibitem [{\citenamefont {Yang}\ \emph {et~al.}(2025)\citenamefont {Yang},
  \citenamefont {Lyu}, \citenamefont {Wang}, \citenamefont {Zhuo},
  \citenamefont {Zhang}, \citenamefont {Wang}, \citenamefont {Shi},
  \citenamefont {Huang}, \citenamefont {Li}, \citenamefont {Song},
  \citenamefont {Li}, \citenamefont {Tong}, \citenamefont {Dou}, \citenamefont
  {Shen}, \citenamefont {Liu}, \citenamefont {Qu},\ and\ \citenamefont
  {Lu}}]{2r8x-9d9m}%
  \BibitemOpen
  \bibfield  {author} {\bibinfo {author} {\bibfnamefont {X.}~\bibnamefont
  {Yang}}, \bibinfo {author} {\bibfnamefont {Z.}~\bibnamefont {Lyu}}, \bibinfo
  {author} {\bibfnamefont {X.}~\bibnamefont {Wang}}, \bibinfo {author}
  {\bibfnamefont {E.}~\bibnamefont {Zhuo}}, \bibinfo {author} {\bibfnamefont
  {Y.}~\bibnamefont {Zhang}}, \bibinfo {author} {\bibfnamefont
  {D.}~\bibnamefont {Wang}}, \bibinfo {author} {\bibfnamefont {Y.}~\bibnamefont
  {Shi}}, \bibinfo {author} {\bibfnamefont {Y.}~\bibnamefont {Huang}}, \bibinfo
  {author} {\bibfnamefont {B.}~\bibnamefont {Li}}, \bibinfo {author}
  {\bibfnamefont {X.}~\bibnamefont {Song}}, \bibinfo {author} {\bibfnamefont
  {P.}~\bibnamefont {Li}}, \bibinfo {author} {\bibfnamefont {B.}~\bibnamefont
  {Tong}}, \bibinfo {author} {\bibfnamefont {Z.}~\bibnamefont {Dou}}, \bibinfo
  {author} {\bibfnamefont {J.}~\bibnamefont {Shen}}, \bibinfo {author}
  {\bibfnamefont {G.}~\bibnamefont {Liu}}, \bibinfo {author} {\bibfnamefont
  {F.}~\bibnamefont {Qu}},\ and\ \bibinfo {author} {\bibfnamefont
  {L.}~\bibnamefont {Lu}},\ }\bibfield  {title} {\bibinfo {title} {Procedure to
  tune a three-site artificial {K}itaev chain to host {M}ajorana bound states
  based on transmon measurements},\ }\href {https://doi.org/10.1103/2r8x-9d9m}
  {\bibfield  {journal} {\bibinfo  {journal} {Phys. Rev. B}\ }\textbf {\bibinfo
  {volume} {112}},\ \bibinfo {pages} {165418} (\bibinfo {year}
  {2025})}\BibitemShut {NoStop}%
\bibitem [{\citenamefont {ten Haaf}\ \emph {et~al.}(2025)\citenamefont {ten
  Haaf}, \citenamefont {Zhang}, \citenamefont {Wang}, \citenamefont {Bordin},
  \citenamefont {Liu}, \citenamefont {Kulesh}, \citenamefont {Sietses},
  \citenamefont {Prosko}, \citenamefont {Xiao}, \citenamefont {Thomas},
  \citenamefont {Manfra}, \citenamefont {Wimmer},\ and\ \citenamefont
  {Goswami}}]{ten_Haaf_2025}%
  \BibitemOpen
  \bibfield  {author} {\bibinfo {author} {\bibfnamefont {S.~L.~D.}\
  \bibnamefont {ten Haaf}}, \bibinfo {author} {\bibfnamefont {Y.}~\bibnamefont
  {Zhang}}, \bibinfo {author} {\bibfnamefont {Q.}~\bibnamefont {Wang}},
  \bibinfo {author} {\bibfnamefont {A.}~\bibnamefont {Bordin}}, \bibinfo
  {author} {\bibfnamefont {C.-X.}\ \bibnamefont {Liu}}, \bibinfo {author}
  {\bibfnamefont {I.}~\bibnamefont {Kulesh}}, \bibinfo {author} {\bibfnamefont
  {V.~P.~M.}\ \bibnamefont {Sietses}}, \bibinfo {author} {\bibfnamefont
  {C.~G.}\ \bibnamefont {Prosko}}, \bibinfo {author} {\bibfnamefont
  {D.}~\bibnamefont {Xiao}}, \bibinfo {author} {\bibfnamefont {C.}~\bibnamefont
  {Thomas}}, \bibinfo {author} {\bibfnamefont {M.~J.}\ \bibnamefont {Manfra}},
  \bibinfo {author} {\bibfnamefont {M.}~\bibnamefont {Wimmer}},\ and\ \bibinfo
  {author} {\bibfnamefont {S.}~\bibnamefont {Goswami}},\ }\bibfield  {title}
  {\bibinfo {title} {Observation of edge and bulk states in a three-site
  {K}itaev chain},\ }\href {https://doi.org/10.1038/s41586-025-08892-5}
  {\bibfield  {journal} {\bibinfo  {journal} {Nature}\ }\textbf {\bibinfo
  {volume} {641}},\ \bibinfo {pages} {890–895} (\bibinfo {year}
  {2025})}\BibitemShut {NoStop}%
\bibitem [{\citenamefont {Bordin}\ \emph {et~al.}(2026)\citenamefont {Bordin},
  \citenamefont {Bennebroek~Evertsz’}, \citenamefont {Roovers}, \citenamefont
  {Torres~Luna}, \citenamefont {Huisman}, \citenamefont {Zatelli},
  \citenamefont {Mazur}, \citenamefont {Ten~Haaf}, \citenamefont {Badawy},
  \citenamefont {Bakkers} \emph {et~al.}}]{bordin2026probing}%
  \BibitemOpen
  \bibfield  {author} {\bibinfo {author} {\bibfnamefont {A.}~\bibnamefont
  {Bordin}}, \bibinfo {author} {\bibfnamefont {F.~J.}\ \bibnamefont
  {Bennebroek~Evertsz’}}, \bibinfo {author} {\bibfnamefont {B.}~\bibnamefont
  {Roovers}}, \bibinfo {author} {\bibfnamefont {J.~D.}\ \bibnamefont
  {Torres~Luna}}, \bibinfo {author} {\bibfnamefont {W.~D.}\ \bibnamefont
  {Huisman}}, \bibinfo {author} {\bibfnamefont {F.}~\bibnamefont {Zatelli}},
  \bibinfo {author} {\bibfnamefont {G.~P.}\ \bibnamefont {Mazur}}, \bibinfo
  {author} {\bibfnamefont {S.~L.}\ \bibnamefont {Ten~Haaf}}, \bibinfo {author}
  {\bibfnamefont {G.}~\bibnamefont {Badawy}}, \bibinfo {author} {\bibfnamefont
  {E.~P.}\ \bibnamefont {Bakkers}}, \emph {et~al.},\ }\bibfield  {title}
  {\bibinfo {title} {Probing {M}ajorana localization of a phase-controlled
  three-site {K}itaev chain with an additional quantum dot},\ }\bibfield
  {journal} {\bibinfo  {journal} {Nat. Commun.}\ }\href
  {https://doi.org/10.1038/s41467-026-68897-0} {10.1038/s41467-026-68897-0}
  (\bibinfo {year} {2026})\BibitemShut {NoStop}%
\bibitem [{Note4()}]{Note4}%
  \BibitemOpen
  \bibinfo {note} {For example, the four-site model has two sweet spot
  conditions, namely $t= \pm \protect \sqrt {\Gamma ^2+\Delta ^2+\protect \frac
  {1}{2} \left (\protect \sqrt {5}+3\right ) \mu ^2}$ and $t= \pm \protect
  \sqrt {\Gamma ^2+\Delta ^2+\protect \frac {1}{2} \left (\protect \sqrt
  {5}-3\right ) \mu ^2}$. Solving Eqs.~\protect \eqref {eq:MP-site}-\protect
  \eqref {eq:MZM-robust} for either of these sweet spots yields expressions for
  $\protect \mathcal {P}^{ab}$ and $\protect \mathcal {R}$ which are
  complicated and do not provide intuitive insights of the behavior of the
  nonlocal MPs.}\BibitemShut {Stop}%
\bibitem [{\citenamefont {Awoga}\ \emph {et~al.}(2017)\citenamefont {Awoga},
  \citenamefont {Bj\"ornson},\ and\ \citenamefont
  {Black-Schaffer}}]{PhysRevB.95.184511}%
  \BibitemOpen
  \bibfield  {author} {\bibinfo {author} {\bibfnamefont {O.~A.}\ \bibnamefont
  {Awoga}}, \bibinfo {author} {\bibfnamefont {K.}~\bibnamefont {Bj\"ornson}},\
  and\ \bibinfo {author} {\bibfnamefont {A.~M.}\ \bibnamefont
  {Black-Schaffer}},\ }\bibfield  {title} {\bibinfo {title} {Disorder
  robustness and protection of {M}ajorana bound states in ferromagnetic chains
  on conventional superconductors},\ }\href
  {https://doi.org/10.1103/PhysRevB.95.184511} {\bibfield  {journal} {\bibinfo
  {journal} {Phys. Rev. B}\ }\textbf {\bibinfo {volume} {95}},\ \bibinfo
  {pages} {184511} (\bibinfo {year} {2017})}\BibitemShut {NoStop}%
\bibitem [{\citenamefont {Awoga}\ \emph {et~al.}(2019)\citenamefont {Awoga},
  \citenamefont {Cayao},\ and\ \citenamefont
  {Black-Schaffer}}]{PhysRevLett.123.117001}%
  \BibitemOpen
  \bibfield  {author} {\bibinfo {author} {\bibfnamefont {O.~A.}\ \bibnamefont
  {Awoga}}, \bibinfo {author} {\bibfnamefont {J.}~\bibnamefont {Cayao}},\ and\
  \bibinfo {author} {\bibfnamefont {A.~M.}\ \bibnamefont {Black-Schaffer}},\
  }\bibfield  {title} {\bibinfo {title} {Supercurrent detection of
  topologically trivial zero-energy states in nanowire junctions},\ }\href
  {https://doi.org/10.1103/PhysRevLett.123.117001} {\bibfield  {journal}
  {\bibinfo  {journal} {Phys. Rev. Lett.}\ }\textbf {\bibinfo {volume} {123}},\
  \bibinfo {pages} {117001} (\bibinfo {year} {2019})}\BibitemShut {NoStop}%
\bibitem [{\citenamefont {Prada}\ \emph {et~al.}(2012)\citenamefont {Prada},
  \citenamefont {San-Jose},\ and\ \citenamefont {Aguado}}]{PhysRevB.86.180503}%
  \BibitemOpen
  \bibfield  {author} {\bibinfo {author} {\bibfnamefont {E.}~\bibnamefont
  {Prada}}, \bibinfo {author} {\bibfnamefont {P.}~\bibnamefont {San-Jose}},\
  and\ \bibinfo {author} {\bibfnamefont {R.}~\bibnamefont {Aguado}},\
  }\bibfield  {title} {\bibinfo {title} {Transport spectroscopy of {NS}
  nanowire junctions with {M}ajorana fermions},\ }\href
  {https://doi.org/10.1103/PhysRevB.86.180503} {\bibfield  {journal} {\bibinfo
  {journal} {Phys. Rev. B}\ }\textbf {\bibinfo {volume} {86}},\ \bibinfo
  {pages} {180503} (\bibinfo {year} {2012})}\BibitemShut {NoStop}%
\bibitem [{\citenamefont {Cayao}\ \emph {et~al.}(2015)\citenamefont {Cayao},
  \citenamefont {Prada}, \citenamefont {San-Jose},\ and\ \citenamefont
  {Aguado}}]{PhysRevB.91.024514}%
  \BibitemOpen
  \bibfield  {author} {\bibinfo {author} {\bibfnamefont {J.}~\bibnamefont
  {Cayao}}, \bibinfo {author} {\bibfnamefont {E.}~\bibnamefont {Prada}},
  \bibinfo {author} {\bibfnamefont {P.}~\bibnamefont {San-Jose}},\ and\
  \bibinfo {author} {\bibfnamefont {R.}~\bibnamefont {Aguado}},\ }\bibfield
  {title} {\bibinfo {title} {Sns junctions in nanowires with spin-orbit
  coupling: Role of confinement and helicity on the subgap spectrum},\ }\href
  {https://doi.org/10.1103/PhysRevB.91.024514} {\bibfield  {journal} {\bibinfo
  {journal} {Phys. Rev. B}\ }\textbf {\bibinfo {volume} {91}},\ \bibinfo
  {pages} {024514} (\bibinfo {year} {2015})}\BibitemShut {NoStop}%
\bibitem [{\citenamefont {Cayao}\ \emph {et~al.}(2017)\citenamefont {Cayao},
  \citenamefont {San-Jose}, \citenamefont {Black-Schaffer}, \citenamefont
  {Aguado},\ and\ \citenamefont {Prada}}]{PhysRevB.96.205425}%
  \BibitemOpen
  \bibfield  {author} {\bibinfo {author} {\bibfnamefont {J.}~\bibnamefont
  {Cayao}}, \bibinfo {author} {\bibfnamefont {P.}~\bibnamefont {San-Jose}},
  \bibinfo {author} {\bibfnamefont {A.~M.}\ \bibnamefont {Black-Schaffer}},
  \bibinfo {author} {\bibfnamefont {R.}~\bibnamefont {Aguado}},\ and\ \bibinfo
  {author} {\bibfnamefont {E.}~\bibnamefont {Prada}},\ }\bibfield  {title}
  {\bibinfo {title} {Majorana splitting from critical currents in {J}osephson
  junctions},\ }\href {https://doi.org/10.1103/PhysRevB.96.205425} {\bibfield
  {journal} {\bibinfo  {journal} {Phys. Rev. B}\ }\textbf {\bibinfo {volume}
  {96}},\ \bibinfo {pages} {205425} (\bibinfo {year} {2017})}\BibitemShut
  {NoStop}%
\bibitem [{\citenamefont {Reeg}\ \emph {et~al.}(2018)\citenamefont {Reeg},
  \citenamefont {Dmytruk}, \citenamefont {Chevallier}, \citenamefont {Loss},\
  and\ \citenamefont {Klinovaja}}]{PhysRevB.98.245407}%
  \BibitemOpen
  \bibfield  {author} {\bibinfo {author} {\bibfnamefont {C.}~\bibnamefont
  {Reeg}}, \bibinfo {author} {\bibfnamefont {O.}~\bibnamefont {Dmytruk}},
  \bibinfo {author} {\bibfnamefont {D.}~\bibnamefont {Chevallier}}, \bibinfo
  {author} {\bibfnamefont {D.}~\bibnamefont {Loss}},\ and\ \bibinfo {author}
  {\bibfnamefont {J.}~\bibnamefont {Klinovaja}},\ }\bibfield  {title} {\bibinfo
  {title} {Zero-energy {A}ndreev bound states from quantum dots in proximitized
  {R}ashba nanowires},\ }\href {https://doi.org/10.1103/PhysRevB.98.245407}
  {\bibfield  {journal} {\bibinfo  {journal} {Phys. Rev. B}\ }\textbf {\bibinfo
  {volume} {98}},\ \bibinfo {pages} {245407} (\bibinfo {year}
  {2018})}\BibitemShut {NoStop}%
\bibitem [{\citenamefont {Cayao}\ and\ \citenamefont
  {Black-Schaffer}(2021)}]{PhysRevB.104.L020501}%
  \BibitemOpen
  \bibfield  {author} {\bibinfo {author} {\bibfnamefont {J.}~\bibnamefont
  {Cayao}}\ and\ \bibinfo {author} {\bibfnamefont {A.~M.}\ \bibnamefont
  {Black-Schaffer}},\ }\bibfield  {title} {\bibinfo {title} {Distinguishing
  trivial and topological zero-energy states in long nanowire junctions},\
  }\href {https://doi.org/10.1103/PhysRevB.104.L020501} {\bibfield  {journal}
  {\bibinfo  {journal} {Phys. Rev. B}\ }\textbf {\bibinfo {volume} {104}},\
  \bibinfo {pages} {L020501} (\bibinfo {year} {2021})}\BibitemShut {NoStop}%
\bibitem [{\citenamefont {Awoga}\ \emph {et~al.}(2022)\citenamefont {Awoga},
  \citenamefont {Cayao},\ and\ \citenamefont
  {Black-Schaffer}}]{PhysRevB.105.144509}%
  \BibitemOpen
  \bibfield  {author} {\bibinfo {author} {\bibfnamefont {O.~A.}\ \bibnamefont
  {Awoga}}, \bibinfo {author} {\bibfnamefont {J.}~\bibnamefont {Cayao}},\ and\
  \bibinfo {author} {\bibfnamefont {A.~M.}\ \bibnamefont {Black-Schaffer}},\
  }\bibfield  {title} {\bibinfo {title} {Robust topological superconductivity
  in weakly coupled nanowire-superconductor hybrid structures},\ }\href
  {https://doi.org/10.1103/PhysRevB.105.144509} {\bibfield  {journal} {\bibinfo
   {journal} {Phys. Rev. B}\ }\textbf {\bibinfo {volume} {105}},\ \bibinfo
  {pages} {144509} (\bibinfo {year} {2022})}\BibitemShut {NoStop}%
\bibitem [{\citenamefont {Awoga}\ \emph {et~al.}(2023)\citenamefont {Awoga},
  \citenamefont {Leijnse}, \citenamefont {Black-Schaffer},\ and\ \citenamefont
  {Cayao}}]{PhysRevB.107.184519}%
  \BibitemOpen
  \bibfield  {author} {\bibinfo {author} {\bibfnamefont {O.~A.}\ \bibnamefont
  {Awoga}}, \bibinfo {author} {\bibfnamefont {M.}~\bibnamefont {Leijnse}},
  \bibinfo {author} {\bibfnamefont {A.~M.}\ \bibnamefont {Black-Schaffer}},\
  and\ \bibinfo {author} {\bibfnamefont {J.}~\bibnamefont {Cayao}},\ }\bibfield
   {title} {\bibinfo {title} {Mitigating disorder-induced zero-energy states in
  weakly coupled superconductor-semiconductor hybrid systems},\ }\href
  {https://doi.org/10.1103/PhysRevB.107.184519} {\bibfield  {journal} {\bibinfo
   {journal} {Phys. Rev. B}\ }\textbf {\bibinfo {volume} {107}},\ \bibinfo
  {pages} {184519} (\bibinfo {year} {2023})}\BibitemShut {NoStop}%
\bibitem [{\citenamefont {Cayao}\ and\ \citenamefont
  {Burset}(2021)}]{PhysRevB.104.134507}%
  \BibitemOpen
  \bibfield  {author} {\bibinfo {author} {\bibfnamefont {J.}~\bibnamefont
  {Cayao}}\ and\ \bibinfo {author} {\bibfnamefont {P.}~\bibnamefont {Burset}},\
  }\bibfield  {title} {\bibinfo {title} {Confinement-induced zero-bias peaks in
  conventional superconductor hybrids},\ }\href
  {https://doi.org/10.1103/PhysRevB.104.134507} {\bibfield  {journal} {\bibinfo
   {journal} {Phys. Rev. B}\ }\textbf {\bibinfo {volume} {104}},\ \bibinfo
  {pages} {134507} (\bibinfo {year} {2021})}\BibitemShut {NoStop}%
\bibitem [{\citenamefont {Hess}\ \emph {et~al.}(2023)\citenamefont {Hess},
  \citenamefont {Legg}, \citenamefont {Loss},\ and\ \citenamefont
  {Klinovaja}}]{PhysRevLett.130.207001}%
  \BibitemOpen
  \bibfield  {author} {\bibinfo {author} {\bibfnamefont {R.}~\bibnamefont
  {Hess}}, \bibinfo {author} {\bibfnamefont {H.~F.}\ \bibnamefont {Legg}},
  \bibinfo {author} {\bibfnamefont {D.}~\bibnamefont {Loss}},\ and\ \bibinfo
  {author} {\bibfnamefont {J.}~\bibnamefont {Klinovaja}},\ }\bibfield  {title}
  {\bibinfo {title} {Trivial {A}ndreev band mimicking topological bulk gap
  reopening in the nonlocal conductance of long {R}ashba nanowires},\ }\href
  {https://doi.org/10.1103/PhysRevLett.130.207001} {\bibfield  {journal}
  {\bibinfo  {journal} {Phys. Rev. Lett.}\ }\textbf {\bibinfo {volume} {130}},\
  \bibinfo {pages} {207001} (\bibinfo {year} {2023})}\BibitemShut {NoStop}%
\bibitem [{\citenamefont {Ahmed}\ \emph {et~al.}(2025)\citenamefont {Ahmed},
  \citenamefont {Tamura}, \citenamefont {Tanaka},\ and\ \citenamefont
  {Cayao}}]{fksg-x8pr}%
  \BibitemOpen
  \bibfield  {author} {\bibinfo {author} {\bibfnamefont {E.}~\bibnamefont
  {Ahmed}}, \bibinfo {author} {\bibfnamefont {S.}~\bibnamefont {Tamura}},
  \bibinfo {author} {\bibfnamefont {Y.}~\bibnamefont {Tanaka}},\ and\ \bibinfo
  {author} {\bibfnamefont {J.}~\bibnamefont {Cayao}},\ }\bibfield  {title}
  {\bibinfo {title} {Odd-frequency pairing due to {M}ajorana and trivial
  {A}ndreev bound states},\ }\href {https://doi.org/10.1103/fksg-x8pr}
  {\bibfield  {journal} {\bibinfo  {journal} {Phys. Rev. B}\ }\textbf {\bibinfo
  {volume} {111}},\ \bibinfo {pages} {224508} (\bibinfo {year}
  {2025})}\BibitemShut {NoStop}%
\bibitem [{\citenamefont {Prada}\ \emph {et~al.}(2020)\citenamefont {Prada},
  \citenamefont {San-Jose}, \citenamefont {de~Moor}, \citenamefont {Geresdi},
  \citenamefont {Lee}, \citenamefont {Klinovaja}, \citenamefont {Loss},
  \citenamefont {Nyg{\aa}rd}, \citenamefont {Aguado},\ and\ \citenamefont
  {Kouwenhoven}}]{prada2019andreev}%
  \BibitemOpen
  \bibfield  {author} {\bibinfo {author} {\bibfnamefont {E.}~\bibnamefont
  {Prada}}, \bibinfo {author} {\bibfnamefont {P.}~\bibnamefont {San-Jose}},
  \bibinfo {author} {\bibfnamefont {M.~W.}\ \bibnamefont {de~Moor}}, \bibinfo
  {author} {\bibfnamefont {A.}~\bibnamefont {Geresdi}}, \bibinfo {author}
  {\bibfnamefont {E.~J.}\ \bibnamefont {Lee}}, \bibinfo {author} {\bibfnamefont
  {J.}~\bibnamefont {Klinovaja}}, \bibinfo {author} {\bibfnamefont
  {D.}~\bibnamefont {Loss}}, \bibinfo {author} {\bibfnamefont {J.}~\bibnamefont
  {Nyg{\aa}rd}}, \bibinfo {author} {\bibfnamefont {R.}~\bibnamefont {Aguado}},\
  and\ \bibinfo {author} {\bibfnamefont {L.~P.}\ \bibnamefont {Kouwenhoven}},\
  }\bibfield  {title} {\bibinfo {title} {From {A}ndreev to {M}ajorana bound
  states in hybrid superconductor--semiconductor nanowires},\ }\href
  {https://doi.org/10.1038/s42254-020-0228-y} {\bibfield  {journal} {\bibinfo
  {journal} {Nat. Rev. Phys.}\ }\textbf {\bibinfo {volume} {2}},\ \bibinfo
  {pages} {575} (\bibinfo {year} {2020})}\BibitemShut {NoStop}%
\bibitem [{Note5()}]{Note5}%
  \BibitemOpen
  \bibinfo {note} {The value numerically jumps to infinity but we have shown it
  as a jump to zero in Fig.~\ref {fig:Schematics}(c,d) for
  clarity.}\BibitemShut {Stop}%
\bibitem [{\citenamefont {Deng}\ \emph {et~al.}(2016)\citenamefont {Deng},
  \citenamefont {Vaitiek\.{e}nas}, \citenamefont {Hansen}, \citenamefont
  {Danon}, \citenamefont {Leijnse}, \citenamefont {Flensberg}, \citenamefont
  {Nyg{\aa}rd}, \citenamefont {Krogstrup},\ and\ \citenamefont
  {Marcus}}]{Deng16}%
  \BibitemOpen
  \bibfield  {author} {\bibinfo {author} {\bibfnamefont {M.~T.}\ \bibnamefont
  {Deng}}, \bibinfo {author} {\bibfnamefont {S.}~\bibnamefont
  {Vaitiek\.{e}nas}}, \bibinfo {author} {\bibfnamefont {E.~B.}\ \bibnamefont
  {Hansen}}, \bibinfo {author} {\bibfnamefont {J.}~\bibnamefont {Danon}},
  \bibinfo {author} {\bibfnamefont {M.}~\bibnamefont {Leijnse}}, \bibinfo
  {author} {\bibfnamefont {K.}~\bibnamefont {Flensberg}}, \bibinfo {author}
  {\bibfnamefont {J.}~\bibnamefont {Nyg{\aa}rd}}, \bibinfo {author}
  {\bibfnamefont {P.}~\bibnamefont {Krogstrup}},\ and\ \bibinfo {author}
  {\bibfnamefont {C.~M.}\ \bibnamefont {Marcus}},\ }\bibfield  {title}
  {\bibinfo {title} {Majorana bound state in a coupled quantum-dot
  hybrid-nanowire system},\ }\href {https://doi.org/10.1126/science.aaf3961}
  {\bibfield  {journal} {\bibinfo  {journal} {Science}\ }\textbf {\bibinfo
  {volume} {354}},\ \bibinfo {pages} {1557} (\bibinfo {year}
  {2016})}\BibitemShut {NoStop}%
\bibitem [{\citenamefont {Sato}\ \emph {et~al.}(2009)\citenamefont {Sato},
  \citenamefont {Takahashi},\ and\ \citenamefont
  {Fujimoto}}]{PhysRevLett.103.020401}%
  \BibitemOpen
  \bibfield  {author} {\bibinfo {author} {\bibfnamefont {M.}~\bibnamefont
  {Sato}}, \bibinfo {author} {\bibfnamefont {Y.}~\bibnamefont {Takahashi}},\
  and\ \bibinfo {author} {\bibfnamefont {S.}~\bibnamefont {Fujimoto}},\
  }\bibfield  {title} {\bibinfo {title} {Non-abelian topological order in
  $s$-wave superfluids of ultracold fermionic atoms},\ }\href
  {https://doi.org/10.1103/PhysRevLett.103.020401} {\bibfield  {journal}
  {\bibinfo  {journal} {Phys. Rev. Lett.}\ }\textbf {\bibinfo {volume} {103}},\
  \bibinfo {pages} {020401} (\bibinfo {year} {2009})}\BibitemShut {NoStop}%
\bibitem [{\citenamefont {Sato}\ \emph {et~al.}(2010)\citenamefont {Sato},
  \citenamefont {Takahashi},\ and\ \citenamefont
  {Fujimoto}}]{PhysRevB.82.134521}%
  \BibitemOpen
  \bibfield  {author} {\bibinfo {author} {\bibfnamefont {M.}~\bibnamefont
  {Sato}}, \bibinfo {author} {\bibfnamefont {Y.}~\bibnamefont {Takahashi}},\
  and\ \bibinfo {author} {\bibfnamefont {S.}~\bibnamefont {Fujimoto}},\
  }\bibfield  {title} {\bibinfo {title} {Non-abelian topological orders and
  {M}ajorana fermions in spin-singlet superconductors},\ }\href
  {https://doi.org/10.1103/PhysRevB.82.134521} {\bibfield  {journal} {\bibinfo
  {journal} {Phys. Rev. B}\ }\textbf {\bibinfo {volume} {82}},\ \bibinfo
  {pages} {134521} (\bibinfo {year} {2010})}\BibitemShut {NoStop}%
\bibitem [{\citenamefont {Oreg}\ \emph {et~al.}(2010)\citenamefont {Oreg},
  \citenamefont {Refael},\ and\ \citenamefont {von
  Oppen}}]{PhysRevLett.105.177002}%
  \BibitemOpen
  \bibfield  {author} {\bibinfo {author} {\bibfnamefont {Y.}~\bibnamefont
  {Oreg}}, \bibinfo {author} {\bibfnamefont {G.}~\bibnamefont {Refael}},\ and\
  \bibinfo {author} {\bibfnamefont {F.}~\bibnamefont {von Oppen}},\ }\bibfield
  {title} {\bibinfo {title} {Helical liquids and {M}ajorana bound states in
  quantum wires},\ }\href {https://doi.org/10.1103/PhysRevLett.105.177002}
  {\bibfield  {journal} {\bibinfo  {journal} {Phys. Rev. Lett.}\ }\textbf
  {\bibinfo {volume} {105}},\ \bibinfo {pages} {177002} (\bibinfo {year}
  {2010})}\BibitemShut {NoStop}%
\bibitem [{\citenamefont {Lutchyn}\ \emph {et~al.}(2010)\citenamefont
  {Lutchyn}, \citenamefont {Sau},\ and\ \citenamefont
  {Das~Sarma}}]{PhysRevLett.105.077001}%
  \BibitemOpen
  \bibfield  {author} {\bibinfo {author} {\bibfnamefont {R.~M.}\ \bibnamefont
  {Lutchyn}}, \bibinfo {author} {\bibfnamefont {J.~D.}\ \bibnamefont {Sau}},\
  and\ \bibinfo {author} {\bibfnamefont {S.}~\bibnamefont {Das~Sarma}},\
  }\bibfield  {title} {\bibinfo {title} {Majorana fermions and a topological
  phase transition in semiconductor-superconductor heterostructures},\ }\href
  {https://doi.org/10.1103/PhysRevLett.105.077001} {\bibfield  {journal}
  {\bibinfo  {journal} {Phys. Rev. Lett.}\ }\textbf {\bibinfo {volume} {105}},\
  \bibinfo {pages} {077001} (\bibinfo {year} {2010})}\BibitemShut {NoStop}%
\bibitem [{\citenamefont {Lutchyn}\ \emph {et~al.}(2018)\citenamefont
  {Lutchyn}, \citenamefont {Bakkers}, \citenamefont {Kouwenhoven},
  \citenamefont {Krogstrup}, \citenamefont {Marcus},\ and\ \citenamefont
  {Oreg}}]{Lutchyn2018Majorana}%
  \BibitemOpen
  \bibfield  {author} {\bibinfo {author} {\bibfnamefont {R.~M.}\ \bibnamefont
  {Lutchyn}}, \bibinfo {author} {\bibfnamefont {E.~P. A.~M.}\ \bibnamefont
  {Bakkers}}, \bibinfo {author} {\bibfnamefont {L.~P.}\ \bibnamefont
  {Kouwenhoven}}, \bibinfo {author} {\bibfnamefont {P.}~\bibnamefont
  {Krogstrup}}, \bibinfo {author} {\bibfnamefont {C.~M.}\ \bibnamefont
  {Marcus}},\ and\ \bibinfo {author} {\bibfnamefont {Y.}~\bibnamefont {Oreg}},\
  }\bibfield  {title} {\bibinfo {title} {Majorana zero modes in
  superconductor-semiconductor heterostructures},\ }\href
  {https://doi.org/10.1038/s41578-018-0003-1} {\bibfield  {journal} {\bibinfo
  {journal} {Nat. Rev. Mater.}\ }\textbf {\bibinfo {volume} {3}},\ \bibinfo
  {pages} {52} (\bibinfo {year} {2018})}\BibitemShut {NoStop}%
\bibitem [{\citenamefont {Wang}\ \emph {et~al.}(2021)\citenamefont {Wang},
  \citenamefont {Xu}, \citenamefont {Li}, \citenamefont {Xu}, \citenamefont
  {Chen},\ and\ \citenamefont {Wang}}]{Wang.2021.Majorana}%
  \BibitemOpen
  \bibfield  {author} {\bibinfo {author} {\bibfnamefont {Z.-H.}\ \bibnamefont
  {Wang}}, \bibinfo {author} {\bibfnamefont {F.}~\bibnamefont {Xu}}, \bibinfo
  {author} {\bibfnamefont {L.}~\bibnamefont {Li}}, \bibinfo {author}
  {\bibfnamefont {D.-H.}\ \bibnamefont {Xu}}, \bibinfo {author} {\bibfnamefont
  {W.-Q.}\ \bibnamefont {Chen}},\ and\ \bibinfo {author} {\bibfnamefont
  {B.}~\bibnamefont {Wang}},\ }\bibfield  {title} {\bibinfo {title} {Majorana
  polarization in non-{H}ermitian topological superconductors},\ }\href
  {https://doi.org/10.1103/PhysRevB.103.134507} {\bibfield  {journal} {\bibinfo
   {journal} {Phys. Rev. B}\ }\textbf {\bibinfo {volume} {103}},\ \bibinfo
  {pages} {134507} (\bibinfo {year} {2021})}\BibitemShut {NoStop}%
\end{thebibliography}%
\end{document}